\documentclass[tradiabstract]{aa} 

\usepackage{graphicx,txfonts}
\usepackage{epstopdf} 
\usepackage{amsmath}
\usepackage{mathtools}
\usepackage{relsize}

\usepackage{natbib,twoopt}
\usepackage[breaklinks=true]{hyperref} 
\bibpunct{(}{)}{;}{a}{}{,} 
\makeatletter
\newcommandtwoopt{\citeads}[3][][]{\href{http://adsabs.harvard.edu/abs/#3}%
{\def\hyper@linkstart##1##2{}%
\let\hyper@linkend\@empty\citealp[#1][#2]{#3}}}
\newcommandtwoopt{\citepads}[3][][]{\href{http://adsabs.harvard.edu/abs/#3}%
{\def\hyper@linkstart##1##2{}%
\let\hyper@linkend\@empty\citep[#1][#2]{#3}}}
\newcommandtwoopt{\citetads}[3][][]{\href{http://adsabs.harvard.edu/abs/#3}%
{\def\hyper@linkstart##1##2{}%
\let\hyper@linkend\@empty\citet[#1][#2]{#3}}}
\newcommandtwoopt{\citeyearads}[3][][]%
{\href{http://adsabs.harvard.edu/abs/#3}
{\def\hyper@linkstart##1##2{}%
\let\hyper@linkend\@empty\citeyear[#1][#2]{#3}}}
\makeatother

\newcommand{\Lsun} {L$_{\odot}$}

\newcommand{\simgreat} {\mathbin{\lower 3pt\hbox{$\rlap{\raise
        5pt\hbox{$\char'076$}}\mathchar"7218$}}}

\newcommand{\simless}{\mathbin{\lower 3pt\hbox {$\rlap{\raise
        5pt\hbox{$\char'074$}}\mathchar"7218$}}}

\begin{document}

\title{Self-consistent 2-phase AGN torus models \thanks{{\it Herschel}
    is an ESA space observatory with science instruments provided by
    European-led Principal Investigator consortia and with important
    participation from NASA.}}  

\subtitle{SED library for observers}

\titlerunning{Self-consistent 2-phase AGN torus models: SED library
  for observers}


%

\author {Ralf~Siebenmorgen\inst{1} \and Frank~Heymann\inst{2}
\and Andreas Efstathiou\inst{3}}

\offprints{Ralf.Siebenmorgen@eso.org}

\institute{European Southern Observatory, Karl-Schwarzschild-Str. 2,
  D-85748 Garching b. M\"unchen, Germany
\and
Deutsches Zentrum f\"ur Luft- und Raumfahrt, Kalkhorstweg 53, 17235 Neustrelitz, Germany
\and
School of Sciences, European University Cyprus, Diogenes Street,
Engomi, 1516, Nicosia, Cyprus }

\date{Received  January xx, 2015 / Accepted xxx, xxx}

\abstract{We assume that dust near active galactic nuclei (AGN) is
  distributed in a torus-like geometry, which may be described by a
  clumpy medium or a homogeneous disk or as a combination of the two (i.e. a 2-phase medium).
  The dust particles considered are fluffy and have higher
  submillimeter emissivities than grains in the diffuse ISM. The
  dust--photon interaction is treated in a fully self-consistent three
  dimensional radiative transfer code.  We provide an AGN library of
  spectral energy distributions (SEDs).  Its purpose is to quickly
  obtain estimates of the basic parameters of the AGN, such as the
  {\it {intrinsic}} luminosity of the central source, the viewing
  angle, the inner radius, the volume filling factor and optical depth
  of the clouds, and the optical depth of the disk midplane, and to
  predict the flux at yet unobserved wavelengths.  The procedure is
  simple and consists of finding an element in the library that
  matches the observations. We discuss the general properties of the
  models and in particular the 10\,$\mu$m silicate band. The AGN
  library accounts well for the observed scatter of the feature
  strengths and wavelengths of the peak emission.  AGN extinction
  curves are discussed and we find that there is no direct one-to-one
  link between the observed extinction and the wavelength dependence
  of the dust cross sections. We show that objects of the library
  cover the observed range of mid IR colors of known AGN. The validity
  of the approach is demonstrated by matching the SEDs of a number of
  representative objects: Four Seyferts and two quasars for which we
  present new Herschel photometry, two radio galaxies, and one
  hyperluminous infrared galaxy. Strikingly, for the five luminous
  objects we find pure AGN models fit the SED without a need to
  postulate starburst activity.}

\keywords{Infrared: galaxies -- Galaxies:
  ISM -- Galaxies: dust}

\maketitle

\section{Introduction}
The suggestion that the 2.2-22\,$\mu m$ emission of Seyfert galaxies
and quasars may be due to dust \citep{LK68} was first explored
analytically by \cite{R69} who were able to reproduce the slope of the
near- to mid IR spectrum. Radiative transfer models of spherically
symmetric distributions of dust in AGN were first developed by
\cite{RC89}, \cite{L91}, and \cite{RRE93}.

Trying to account for the infrared emission of type 1 and type 2 AGNs
with models that assume a disc-like or toroidal distribution of dust
was recognized early on as an important test of the unified model for
AGN \citep{A93}. Such models have been developed from the early 1990s
\citep{ERR91, PK92, PK93, RR93, GD94, ERR95,E95}. Most of the
discussion on these models concerned the interpretation of the
observed behaviour of the silicate feature at 9.7\,$\mu m$ which was
absent in emission in type 1 objects \citep{R91} and challenged the
whole idea of trying to explain the infrared emission of AGN with dust
models and the unified model itself. It is now recognized that smooth
and clumpy torus models \citep{F12} can reproduce the whole range of
features which are observed using Spitzer spectroscopy
(\citealp{S05,H05,Spoon07,Schweitzer08,MC15,Hatz15}) as well as high
spatial resolution ground based data (\citealp{H10, AH11,
  GonzalesMartin13,Esquej14,Ramos14}).  Another challenge for these
models is trying to explain the spatial extent of the mid IR emission
which also looked puzzling initially. A significant part of the mid IR
emission appears to come from the ionization cones \citep{B93, C93,
  H12,Hoenig13}, which suggests the presence of dust in the ionization
cones either in the form of discrete clouds or as outflows \citep{E95,
  Elitzur06}.  The data suggest that what is needed might be a
combination of smooth and clumpy structures in AGN which may extend
into the narrow line region, see \cite{Netzer15} for a recent review.
The present picture is also supported by recent hydrodynamical
simulations which predict a filamentary torus and a turbulent disc
component \citep{Sch14}.

Large ground-based telescopes offer an opportunity to much better
isolate the contribution of the host galaxy and circumnuclear star
formation from the AGN when compared to observations from space
(\citealp{AH11,Ramos11,Asmus14}).  Spitzer spectra provide a
resolution of $\sim 3$'' whereas high resolution MIR spectra from the
ground are an order of magnitude sharper, and sensitive enough to
observe local Seyferts at luminosities $<10^{46}$ erg/s.  Such high
resolution studies have been undertaken for example in NGC~1068,
NGC~1365, NGC~3281, and Circinus (\citealp{Mason12, Roche06, AH12,
  Sales11, Esquej14}). A sample of 19 Seyferts are presented by
\cite{H10} and another 22 Seyferts are studied by
\cite{GonzalesMartin13}. At $\sim 20$\,pc resolution
\cite{Ruschl-Dutra14} detected the silicate feature at 9.7\,$\mu$m in
absorption for the Seyfert 2 NGC~1386, and in emission for the Seyfert
1 NGC~7213 with a peak wavelength beyond 10.5\,$\mu$m. \cite{Ramos14}
compared the nuclear MIR spectrum with the Spitzer spectrum and find
that the AGN in Mrk1066 dominates the continuum emission at $\lambda <
15$\,$\mu$m on scales of $\sim 60$\,pc (90 – 100\%) whereas this
contribution decreases to below 50\% when the emission of the central
$\sim 830$\,pc is considered.

IR interferometry offers unprecedented angular resolution that allows
resolving the inner regions of dusty AGN (\citealp{Jaffe04, T07,
  T09}). It has been shown that the inner torus radii derived from
visibility curves as expected scale approximately with the square root
of the AGN luminosity (\citealp{Suganuma06, Kishi11}), except for some
deviations that are observed for fainter sources \citep{Burtscher13}.
A sample of nearby Seyfert galaxies observed with IR interferometers
\citep{Hoenig13} reveal a diversity of obscuring structures. A
significant part of the mid IR emission of certain sources appears to
come from the the polar direction. These observations are difficult to
reconcile with torus models alone and challenges the justification of
using clumpy models to fit that emission \citep{Hoenig13}. Polar dust
may be physically disconnected from the torus and may arise from a
radiatively driven dusty wind blown off the inner region of the torus.

ALMA observations have recently been reported for the Seyferts
NGC~1566 \citep{Combes14}, NGC~1068 \citep{GarciaBurillo14} and NGC~34
\citep{Xu14}.  Typically at 25-50pc resolution an unresolved core and
a circumnuclear disk extending about 200pc in radius is detected.

Solutions of the radiative transfer in clumpy AGN tori employ methods
that have different levels of sophistication.  A first formalism was
presented by \cite{Nenkova02} utilizing a 1d radiative transfer code
and computed SED of clouds by assuming a slab geometry.  The slabs are
heated from different angles and then averages are taken so that they
mimic the emission of spheres. These spheres are arranged so that a
torus-like geometry is simulated. \cite{Hoenig06} pre-computed these
clouds in 2d and \cite{Hoenig10} with an upgrade of that model in
3d. In the AGN torus there are clouds that are not illuminated
directly by the central source because other optically thick clouds
lie along the line of sight towards the primary heating source. Such
clouds are shadowed and indirectly heated. In the above mentioned
models it is assumed that this indirect heating is coming from clouds
in the immediate surrounding which are directly heated so that
averages can be build.  Clumps, which are directly heated by photons
reflected on a surface of an optically thick disk were not
treated. Clumpy torus models are successful in explaining the
strengths of the silicate emission and absorption band. {\bf
  {However,}} they do not account for very deep silicate absorptions
and under-predict the NIR emission \citep{Nenkova08, Spoon07}.
Self-consistent 3d clumpy torus models are presented by \cite{Sch08}
and \cite{Sta12}.  \cite{MC15} find that Seyfert galaxies have clumpy
dust distributions and/or a disk component. The formation of a clumpy
and a continuous dust component is also predicted by hydrodynamical
simulations \citep{Sch14}.

In this paper we describe our SED library of 2-phase AGN torus models
that has been computed with the fully self-consistent 3d Monte Carlo
radiative transfer code of \cite{HS12}.  We apply beside the clumps a
density distribution of a disk component that extends into the
ionization cones. We will show below that the emission from such
diffuse dust in the ionization cones can dominate the total emission
from an edge-on view. This can explain why some Seyferts show extended
emission coincident with the ionization cones at 10$\mu$m despite the
fact that there is an optically thick torus which lies in a plane
perpendicular to the ionization cones. The emission coming from the
ionization cones is also much more compact than the torus in agreement
with interferometric observations {\bf {\citep{Burtscher13,
      Hoenig13}}}.  We discuss the general properties of the models
(Sect.~3) and apply the SED library to fit a number of representative
objects (Sect.~4). We summarize our findings in the conclusions
(Sect.5).

\begin{figure}[htb]
  \includegraphics[width=9cm]{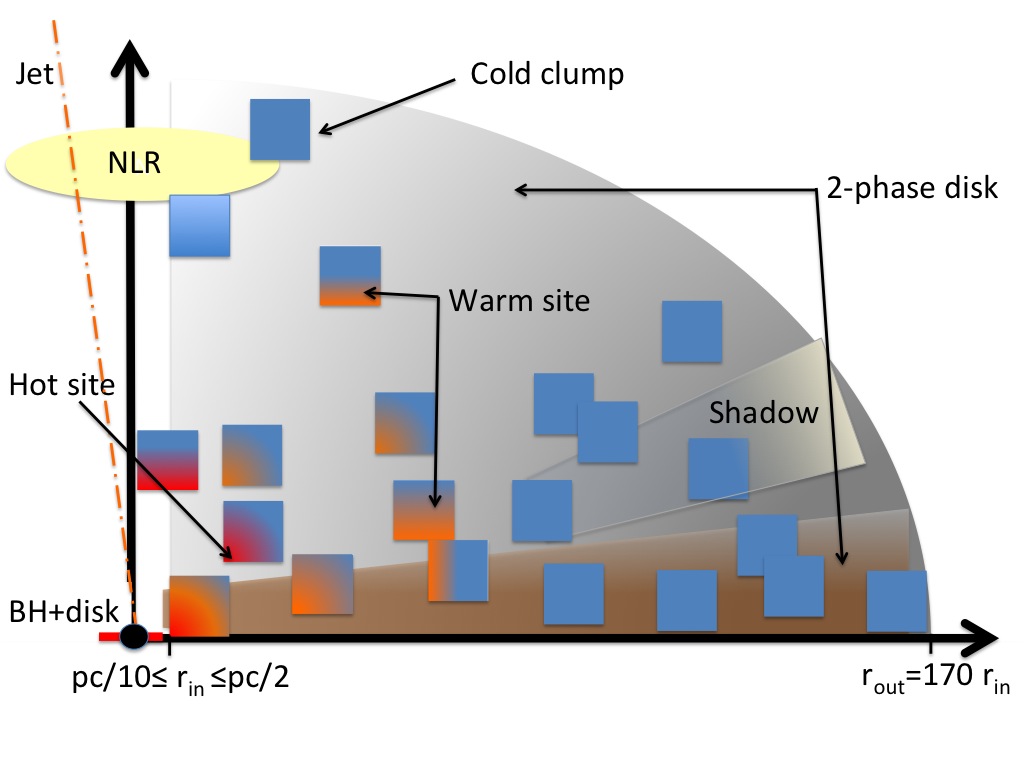}
  \caption {Schematic view of the AGN structure assumed in this
    work. Dust extends from the dust evaporation radius at $pc/10 \leq
    r_{\rm {in}} \leq pc/2$ up to distance of $ r_{\rm {out}} = 170
    \times r_{\rm {in}}$ from the black--hole (BH). Note that dust
    from the smooth disk emerges also as a diffuse component into the
    polar region.  Dust clouds are pc sized with cube length of $d = 3
    \times r_{\rm {in}}$. The hot, warm and cold side of the clouds is
    indicated and for one clump their shadows.  The accretion disk in
    the immediate black--hole environment and the jet is not
    treated. \label{model.fig}}
\end{figure}

\section{AGN  Model}

In this section we describe the AGN model library of SEDs. We outline
the radiation transfer code, the assumed dust structure, and the
library set-up. We choose a condensed description of the AGN model
with all parameters kept constant except a minimum of five key
parameters that are varied to compute the library.  We visualize the
dust density distribution and AGN images at different wavelengths, and
show the impact of our free parameters on the SED.

Most models for the torus in AGN consider dust that is representative
for the diffuse interstellar medium and the fact that there are huge
differences in the density and the radiation environment of both media
is ignored. Our library is computed considering fluffy grains. The
rational for this and the dust properties assumed are discussed in
Sect.~\ref{dust.sec}. Finally, we discuss the impact of scattering in
spatially unresolved observations of the extinction curve in AGN.

\subsection{Radiative Transfer}

For our calculations we apply the Monte Carlo (MC) continuum radiative
transfer code of \cite{HS12}.  The code allows us to follow the light
path of photons that are emitted from a heating source through an
arbitrary three dimensional (3d) dust configuration.  Photons are
binned into packets of equal energy, those photons of one packet have
the same frequency.  Dust absorption, scattering and emission events
are computed by chance. The process depends on the optical depth and
the absorption and scattering cross section of the dust.  We utilize
the MC algorithm developed by \cite{BW01}, which is iterative-free as
it employs an immediate update of the dust temperatures. For very
optically thin regions we use the method by \cite{L99} which reduces
uncertainties in the dust temperatures. Anisotropic scattering is
treated assuming the phase function $p(\cos(\theta))$ as given by
\citep{HG41}

\begin{equation} 
p(\cos(\theta)) = \frac{1}{4\pi}\, \frac{1-g^2}{(1 + g^2 - 2 g \,
\cos(\theta))^{3/2}} \,.
\end{equation} 

\noindent 
where $\theta$ is the scattering angle. The anisotropy factor $g$ of
the grains is provided by the dust model and whenever $g=0$ we
consider isotropic scattering.

The model space is set up as cubes in an orthogonal Cartesian
grid. Each cube can be divided into an arbitrary number of sub-cubes
with constant density.  An illustration of the grid is given in Fig.1
by \cite{HS12}. The subdivision of cubes allows a finer sampling
whenever required. The advantage of such a grid is that it leads to a
huge reduction of the computer memory for some applications.  Examples
of this are the interface of optically thick clumps that are embedded
in an otherwise homogeneous medium, regions close to the dust
evaporation zone or configurations where the ratio of the outer and
inner radius from the central source where dust exists is large
$r_{\rm {out}} / r_{\rm {in}} > 100$. The number of cells can be
reduced further by a factor of eight by locating the source at origin
and assuming mirror--symmetry with respect to the planes at $x = 0$,
$y = 0$ and $z = 0$. Vectoring the code so that it can run on shared
memory devices reduces the huge computing time of the MC
procedure. The number of photon packets that can be followed in
parallel on such machines scales roughly with the number of processing
units available and reduces the required computing time accordingly.

Ray-tracing in parallel projection from the detector of the observer
through the model space, using temperatures and scattering events for
each cell of the MC run, allows the generation of high spatial
resolution and high signal-to-noise images. In this work we derive the
SED of the object by counting the photon packets that eventually
escape the model space into a particular solid angle. This method
gives within the photon noise the same results when compared to SEDs
computed by ray tracing.  The later method has the advantage that it
provides images for each frequency. However, it is expensive in
computing time. For one AGN model and SEDs computed in 9 viewing
direction the ray tracer takes about half a day of computing time
while the photon counting is completed in a few seconds.

\subsection{AGN Structure \label{agnstruct}}

The super-massive central black hole with its accretion disk
\citep{Shakura} is considered to be the primary heating source of the
circumnuclear environment. It emits anisotropic hard UV and optical
photons \citep{N87, KM11}.  The anisotropy of the primary source does
not have a strong impact on the torus spectrum when compared to models
with isotropic heating. The overall SED shape and the 10\,$\mu$m Si--O
dust feature are not seriously affected by the anisotropy
\citep{Sta12}. Therefore the emission of the primary heating source is
taken to be isotropic. In optically thick applications the spectral
shape of the heating source has virtually no impact on the dust
emission spectrum. We assume that the primary energy source takes a
spectral shape approximated by \cite{RR95}:

\begin{equation} \label{agn.eq}
\lambda F_\lambda \propto 
\left \{ \begin{array}{ccllll} 
\lambda^{1.2} & \text{for} & ~-\infty ~&< &\lambda ~< 0.01 & \rm{(}\mu\rm{m)} \\  
\ {\rm {const}} & \text{for} & ~~0.01 &<~ &\lambda ~< 0.1 & \rm{(}\mu\rm{m)} \\  
\lambda^{-0.5} & \text{for} & ~~0.1~~ &<~ &\lambda ~< ~~1 & \rm{(}\mu\rm{m)} \\  
\lambda^{-3} & \text{for} & ~~1.0~~ &<~  &\lambda ~< +\infty & \rm{(}\mu\rm{m)} \\ 
\end{array} \right. 
\end{equation} 

\noindent

The inner radius $r_{\rm{in}}$ of the dust torus is set close to the
dust evaporation region, which scales with the total AGN luminosity as
$L_{\rm {AGN}} \propto r_{\rm{in}}^{2}$. For type~1 AGN this relation
is confirmed by reverberation measurements and interferometry
\citep{Kishi11}.  As long as this relation is respected, the SED
becomes scale invariant, so that the SED shape does not vary for
different AGN luminosities.  We therefore keep the luminosity of the
primary source constant at $L_{\rm {11}} = 10^{11}$\Lsun \/. The
sublimation temperature of the dust depends on details such as grain
mineralogy, porosity and size. We assume that it is somewhere between
800 -- 1800\,K and use $r_{\rm{in}}$ as a free-parameter.

The required maximum size of the torus depends on the observed
wavelength. However, choosing $r_{\rm{out}}$ smaller than the size
corresponding to the temperature, respectively wavelength, of interest
results in an artificial cut-off in the brightness distribution of the
models.  For example, when selecting
$r_{\rm{out}} /r_{\rm{in}} \sim 25$ the cut-off occurs at temperatures
of $\sim 400$\,K, which corresponds to wavelengths around
8\,$\mu$m. It is therefore important to select $r_{\rm{out}}$ large
enough for all cases to avoid artificial cut-off in the brightness
distribution \citep{Hoenig10}.  We assume that the dust density in the
torus declines with distance from the AGN, so that at some point a
further increase of $r_{\rm{out}}$ has no impact on the SED. We find
$r_{\rm{out}} /r_{\rm{in}} \sim 170$ is a reasonable choice. This is
larger than that assumed in most of the previous AGN models.

The distribution of the dust in AGN is still a matter of debate.  It
was realized early on that the distribution of gas and dust is most
likely clumpy and in a torus-like structure \citep{KB88}. To account
for IR observations and the 10\,$\mu$m silicate feature radiative
transfer calculations of clumpy dust distributions have been carried
out, e.g. \citep{Nenkova02, Sch08, HS12, Sta12}. Dust models that assume
homogeneous or smooth distributions \citep{PK92, ERR95, E14} have also
been considered. AGN models assuming clumpy or smooth dust
distributions can produce similar SEDs \citep{F12}.

Interferometric observations of nearby AGN have so far resolved
disk-like or extended dust emission components (usually coincident
with the ionization cones) down to pc scale. Unresolved structures
such as clumps are below the accessible resolution limits of $\sim
15$\,mas \citep{T14}. X--ray monitoring of discrete absorption eclipse
events of Seyferts is presented by \cite{M14}.  They find a higher
detection probability of clumps for type 2 than for type 1, and from
the observed column density an optical depth of the clumps between
$\tau_{\rm V} = 20 - 160${\footnote{Unless otherwise stated the
  optical depth is specified for the V band. }} is estimated. The current
data of the X--ray study provides insight into the cloud distribution
only close to the dust evaporation zone. For type~2 objects the X--ray
absorption can also be explained by a homogeneous medium.

We consider a dust structure in the AGN which is approximated by an
isothermal disk that is embedded in a clumpy medium. The midplane of
the disk is at $z=0$, so in the xy-plane of the coordinate system the
dust density distribution of the disk is given by:

\begin{equation}
\rho(x,y,z) = \rho_{o} \, \frac{r_{\rm {out}}}{r} \,   \rm{e}^{- {\pi}
  \, ( \frac{z} {2 \,h})^2}.
\end{equation}

\noindent
 
We choose $ \rho_{o}$ so that the optical depth of the disk midplane
in the V band, which is measured between $r_{\rm {in}}$ and $r_{\rm
  {out}}$, is $\tau_{\rm {V, mid}} = 0, 30, 100, 300$ or 1000.  The scale
height is approximated by \citep{P04}:

\begin{equation}
h =\frac{r_{out}}{8} \left( \frac{2r}{r_{\rm {out}}} \right)^{1.125}  .
\end{equation}

In addition there are a number $N_{\rm {cl}}$ of clouds that are
passively heated from outside. The possibility that massive stars heat
clouds from inside is not treated. The density of the clouds is
assumed to be constant. This is consistent with the small scatter in
the column density of the clouds as derived from X-ray observations
\citep{M14}. The clouds are arbitrarily distributed in the model
space. We assume, for random numbers denoted by $\mathfrak{z}$, that
the clouds have an isotropic distribution along the azimuth angle
$\phi = \pi/2 \, \mathfrak{z}$, but they are weighted along the polar
angle $\theta$ by $\cos(\theta) = \mathfrak{z}^2$, so that there are
more clumps at low than at high latitudes, and along the radial
direction $r$ an equidistant separation with the disc midplane given
by:

\begin{equation}
\label {eq.cloudist}
  r = r_{\rm {in}} \, + \, \sqrt{\mathfrak{z}(r_{\rm {out}}^{2}\,
    -\, r_{\rm {in}}^{2})}
\end{equation}

A schematic of the AGN model is shown in Fig.~\ref{model.fig}.  The
potential accretion disk in the immediate black--hole environment,
where dust cannot survive, and the jet emission is not treated in our
model.  The dust disk extends from a radius of less than 1 up to a few
hundred pc distance from the primary heating source. We assume dust
clouds have a cube size of 0.3 -- 3\,pc. The hot, warm and cold side
of these clumps and the shadowing caused by them is computed
self--consistently in the 3d radiative transfer model. Please note
that all scales correspond to an AGN luminosity of $10^{11} L_\odot$.
As discussed above all radii scale as $L^{1/2}$.

\subsection{Technical set-up}

The model grid is set up into $(51,51,51)$ basic cubes with a side
length of a cube $d$ depending on the choice of the inner dust
radius. It is given by $d = 3.\bar{3} \, r_{\rm {in}} $. The central
cube has $(900, 900, 900)$ subcubes, and along each direction the
following $(10, 10, 10)$ basic cubes are each subdivided into $(15,
15, 15)$ subcubes.  We can achieve a resolution of $4 \times
10^{14}$\,cm in the inner part of the torus, whereas by comparison the
resolution in the hydrodynamical simulations of \cite{Sch14} was $4
\times 10^{17}$\,cm. Each cloud is divided into (3, 3, 3) basic cubes
and each of these is divided into $(15, 15, 15)$ sub-cubes. This gives
a total of more than 90,000 cells for each clump, which ensures that
even for the highest optical depth cases each sub-cube of the clump
has an optical depth along its axis of $\tau_{\rm V} \simless 1$. This
criterion is similar to the one used by \cite{Hoenig06} who
demonstrated that the cell temperatures converge in that case (see
their Table 1).  \cite{Sta12} used clouds with an optical depth
$\tau_{\rm V} \simgreat 100$. These clumps are divided into $(8, 8,
8)$ cells when the cloud number is low whereas only a single cell is
used when the number of clouds is high.  The latter simplification is
not justified whenever the optical depth of a sub-cube of a cloud
becomes larger than 1. SEDs computed for optically thick clouds and
ignoring sufficient sub-sampling drastically underestimate the IR
emission.

The AGN spectrum (Eq.~\ref{agn.eq}) is divided into 256 frequency bins
and for each bin 200,000 photon packets are emitted. In our treatment
this is sufficient to achieve a reasonable signal--to--noise of fluxes
in the submillimeter.  SEDs are computed in different viewing angles
that are equally spaced in $\cos \theta$ between $0^{\rm o}$ and
$90^{\rm o}$. The SEDs are reasonably well sampled by a division into
nine different viewing angles.

\begin{figure}

\includegraphics[scale=0.37,angle=0.9,clip=true,trim=1cm 0cm 0cm 1cm]{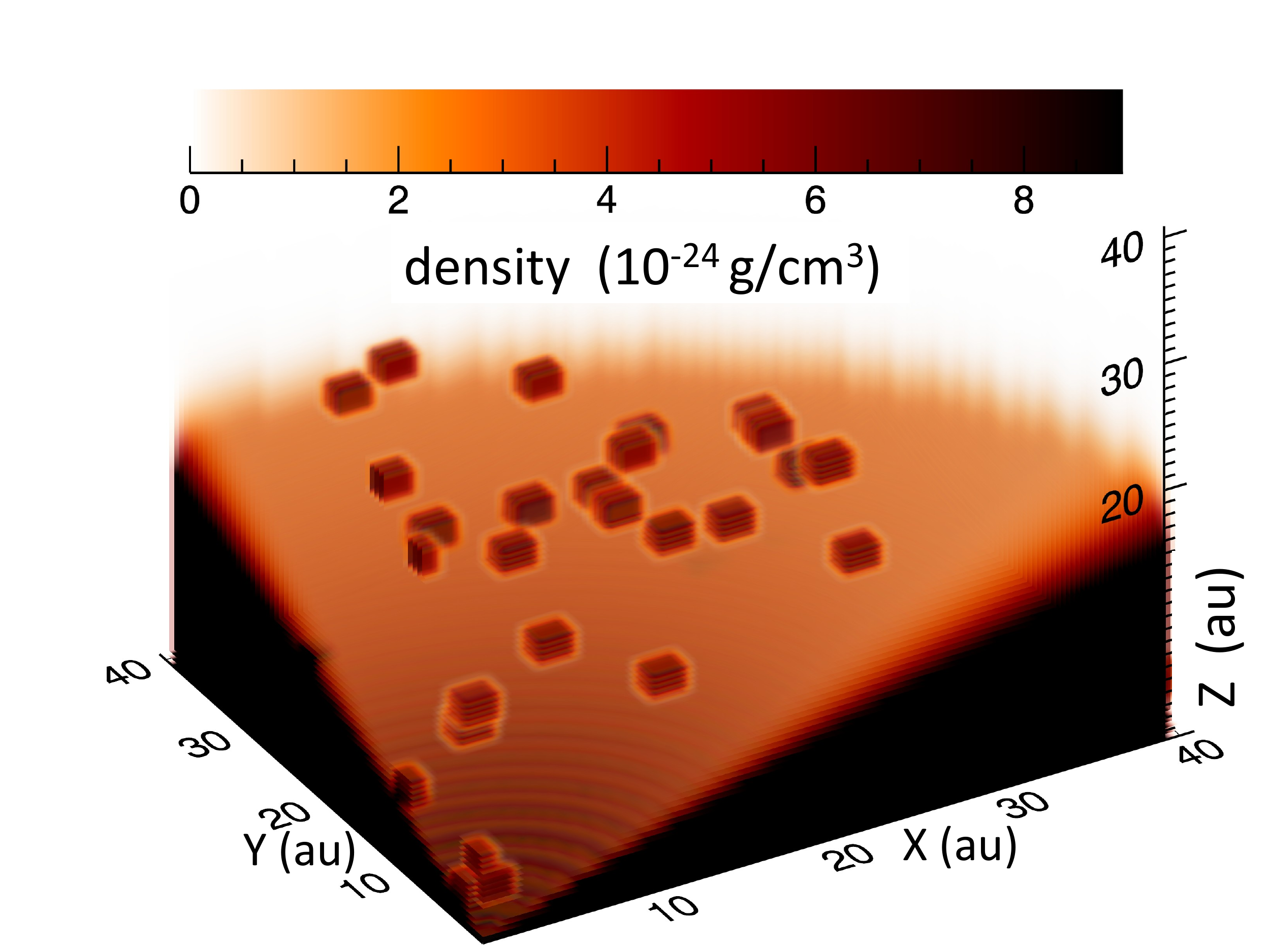}

\includegraphics[scale=0.36,angle=0.93,clip=true,trim=0cm 0cm 0cm 1cm]{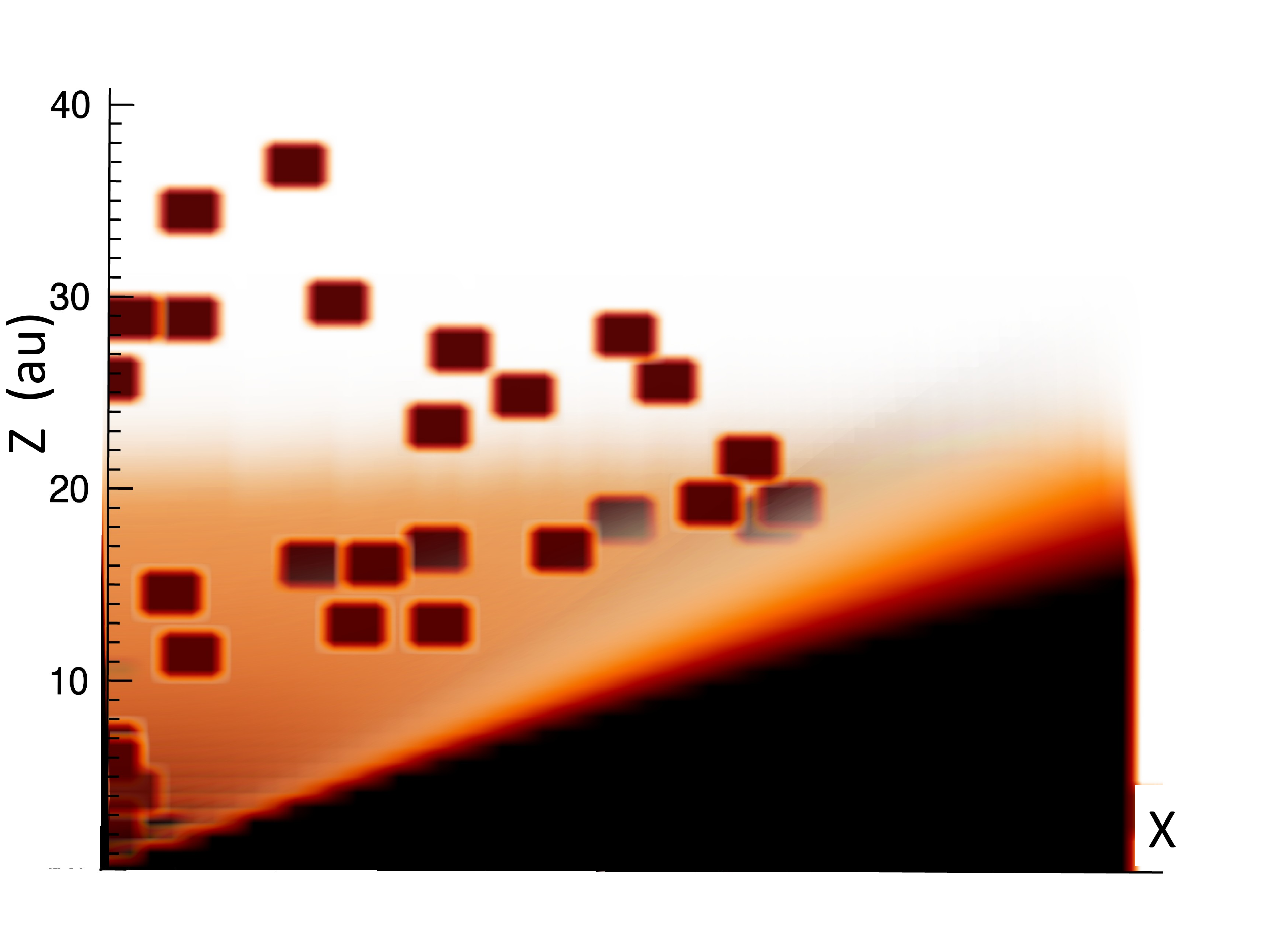}

\caption{Visualization of the dust density distribution of an octant
  of the AGN model space. A view from high latitudes (top) and
  edge--on along the x,z axis (bottom) is shown. Clumps deep inside
  the disk are not visible.  Note the presence of clumps in the polar
    region. The axes are in arbitrary units. \label{3d.pdf} }
\end{figure}


\subsection {Library setup}

We find that for seeting up a SED library of AGN there are in our
prescription besides the viewing angle four key parameters: (1) the
inner radius $r_{\rm {in}}$, (2) the volume filling factor of the
clouds $\eta$, (3) the optical depth of the clouds $\tau_{\rm {V,
    cl}}$, and (4) the optical depth of the disk midplane $\tau_{\rm
  {V, mid}}$. There are other parameters, such as the outer radius,
$r_{\rm{out}} = 170 \times r_{\rm{in}}$, up to where dust exists,
details of the clump structure (Sect.~\ref{agnstruct}) or their
distribution function (Eq.~\ref{eq.cloudist}), which are all kept
constant. The variations of the free parameters of the SED library are
summarized in Table~\ref{mod.tab}. In the AGN library there are 400
models each providing SEDs for nine viewing directions.

We varied the density profile within a cloud, e.g. following a $1/r$
profile, and experimented with the cloud structure, for example using
spherical clumps instead of cubes. We confirm the results of
\cite{Hoenig10}, who concluded that unless very peculiar clump
surfaces are used, the exact shape of the clumps does not matter too
much and a spherical or cubic representation probably catches the
essence of the dust clouds in AGN tori. These authors find somewhat
redder MIR colors when distributing more dust clouds at larger
distances. We also varied the cloud distribution along the radial
direction and the sizes of the clouds. We also experimented with using
a large number of small clouds in the inner region and a small number
of large clouds in the outer region.  We compared models of the same
total cloud mass and found that all of these changes of the cloud
properties and distribution functions has some but no significant
impact on the SED. We therefore do not consider variation of these
properties in the library.

Finally, we implemented also clumps with parameters as derived from
X--ray eclipse observations \citep{M14}. Such clumps are detected in
the central region of the AGN, so in our model in the innermost
cube. They are assumed to be as small as $3.\bar{3} \times
10^{14}$\,cm and have $\tau_{\rm V} = 20 - 160$. For these X--ray
clumps we computed a single clump temperature. We find that clumps
detected by X-ray eclipse events have marginal impact on the SED as
long as their number stays below several million. Therefore we neglect
such X--ray clumps in the rest of the paper.


\begin{table}[htb]
\begin{center}
\caption {Parameters of the AGN library. \label{mod.tab}}
\begin{tabular}{|l|l|l|}
 \hline 
& & \\
Parameter & Symbol & Values \\ 
& & \\
\hline
Viewing angle$^{a)}$ & $\theta$ ($^o$) & 86, 80, 73, 67, 60,  \\ 
              &  & 52, 43, 33, 19\\ \hline
Inner radius$^{b)}$ & $r_{in}$ ($10^{17}$cm) & 3, 5.1, 7.7, 10, 15.5 \\   \hline 
Cloud volume & $\eta$  (\%) & 1.5, 7.7, 38.5, 77.7 \\ 
filling factor$^{c)}$  &         &  \\  \hline 
Cloud optical depth$^{d)}$  & $\tau_{\rm {V, cl}}$ & 0, 4.5, 13.5, 45 \\ 
\hline 
Optical depth of    & $\tau_{\rm {V, mid}}$ & 0, 30, 100, 300, 1000 \\ 
disk midplane$^{e)}$& & \\ \hline 
\end{tabular}
\end{center} {\bf {Notes.}} $^{a)}$ The viewing angle is equally
binned in $\cos(\theta)$ and measured from the z--axis to the x,y
plane.  $^{b)}$ The luminosity of the primary source is kept constant
at $L_{\rm {11}} = 10^{11}$\Lsun \/. $^{c)}$ The cloud volume
fraction corresponds to the number of clouds of $N_{\rm {cl}} = 160$,
800, 4000, 8000 within the 3d model space.  $^{d)}$ The cloud optical
depth is measured along the edge of the cloud that has a structure of
a cube. $^{e)}$ The disk midplane is located in the x,y-plane at
$z=0$. We measure $\tau_{\rm {V, mid}}$ along the midplane from the
source to the outer edge of the disk.

\end{table}


\begin{figure*}
\begin{center}
 \includegraphics[scale=0.6]{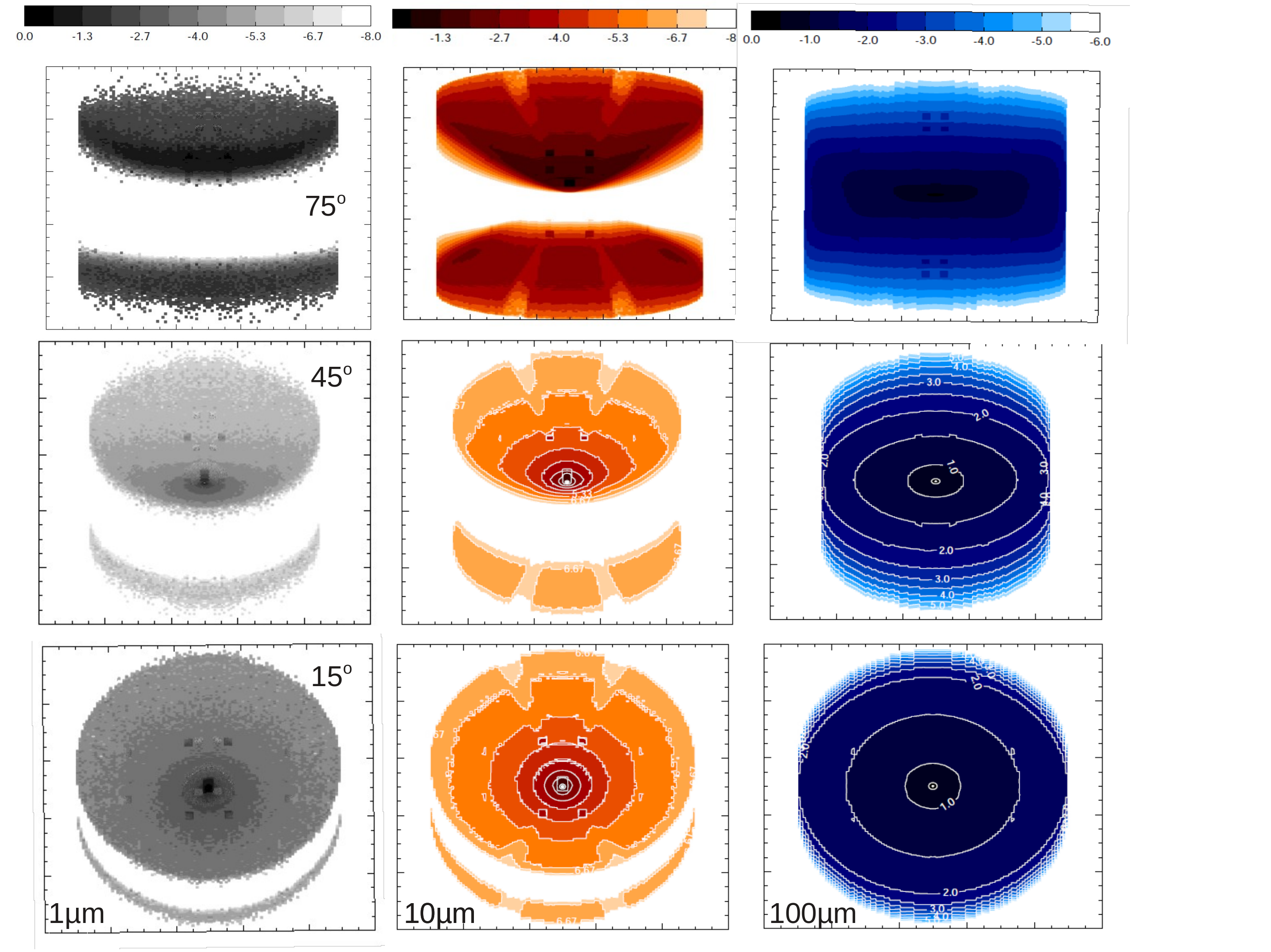}
 \caption{Images of the AGN torus computed by ray-tracing at viewing
   directions corresponding to 15$^{\rm{o}}$ (face--on, bottom),
   45$^{\rm{o}}$ (middle) and 75$^{\rm{o}}$ (edge--on, top) and from
   left to right at wavelengths of 1 (grey), 10 (orange), and
   100\,$\mu$m (blue).  The images are logarithmically scaled and at
   each wavelength the flux is normalised to its peak
   value. \label{images.pdf}}
\end{center}
\end{figure*}

\subsection {Visualization}

As an aid in visualizing the AGN structure we show the dust density
distribution and images at different wavelengths. The particular
parameters of the AGN model we visualize are $r_{\rm {in}} =
0.25$\,pc, $\tau_{\rm {V, cl}} = 4.5$, $\tau_{\rm {V, mid}}=1000$, $\eta =
1.5$\,\%.  A 3d visualization of the dust density distribution for an
octant of the model space as seen from high latitudes and in edge--on
view is shown in Fig.~\ref{3d.pdf}.  The density inside the disk is so
high that clumps are only visible above or within the upper surface
layer of the disk.

Images of the same AGN model are shown in Fig.~\ref{images.pdf} for
different viewing directions and wavelengths. The image at 1\,$\mu$m
is dominated by scattered light with an exception in the face--on view
where a small contribution by dust emission is visible in the
innermost region.  In scattered light only the upper and lower
surfaces of the disk are visible. The optical depth is too high for
the photons to penetrate deeper into the disk. This explains why there
are two distinct gray areas noticeable. Clumps are visible as the few
bright dots in the middle and bottom panel (left). The image at
10\,$\mu$m displays emission by warm dust and scattering is no longer
important. The optical depth becomes smaller and one probes deeper
inside the disk (compare top--left and top--middle panel of
Fig.~\ref{images.pdf}). The disk itself remains optically thick at
10\,$\mu$m so that two distinct surfaces are visible as in the
scattering light images. As can be seen in the top-middle panel, the
emission from the diffuse dust in the ionization cones can dominate
the total emission from an edge-on view. This can explain why some
Seyferts show extended emission coincident with the ionization cones
at 10$\mu$m despite the fact that there is an optically thick torus
which lies in a plane perpendicular to the ionization cones. The image
at 10\,$\mu$m shows also the shadowing of the clumps. They can be
noticed as notched contours or jet--like structures (top--middle).
Behind a clump there is less direct heating from the central source,
the dust becomes colder and the 10\,$\mu$m flux is reduced. The image
at 100\,$\mu$m reveals the cold dust emission and at this wavelength
the disk is optically thin and the radiation appears isotropic.

\begin{figure}
\centering \includegraphics[scale=0.62,clip=true,trim=3.2cm 6.1cm 3.1cm
  6cm]{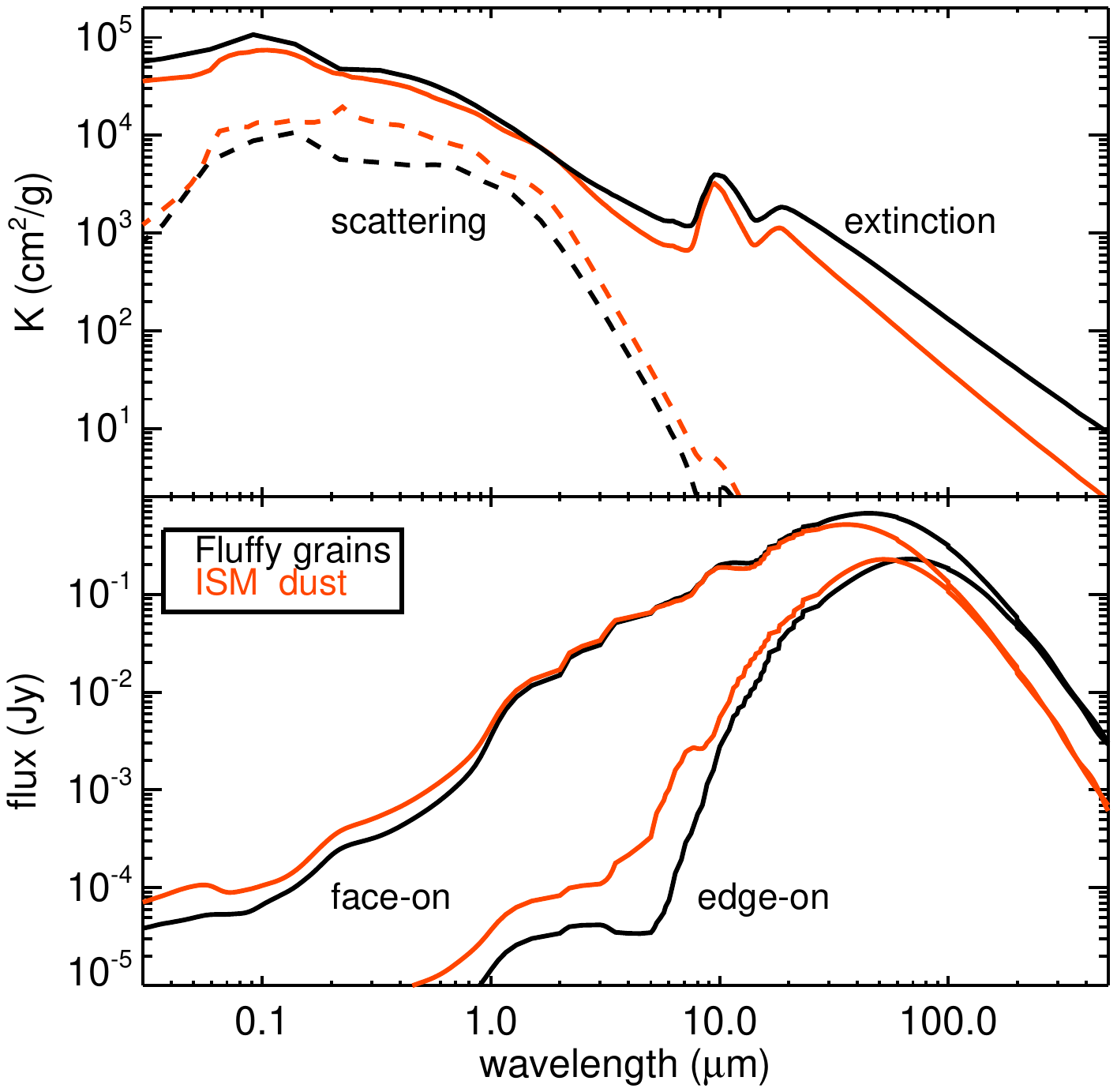}
  \caption{Extinction (full line) and scattering (dashed) cross
    section $K$ as a function of wavelength ({\it {top}}).  We show
    fluffy aggregates (black) of silicates and amorphous carbon
    \citep{KS94} and identical materials as unmixed bulk particles
    (red) of the diffuse ISM \citep{S14}. For both dust models and
    with otherwise identical parameters the emission spectrum of a
    clumpy AGN torus is given for face-on and edge--on view ({\it
      {bottom}}). \label{FluffyISM.pdf} }
\end{figure}

The size of the torus becomes wavelength dependent if one cuts the
image contrast at higher flux levels, e.g. take $\log (F /F_{\rm
  {max}}) > - 4$ in Fig.~\ref{images.pdf} \citep{Sta12b}.  At shorter
wavelengths radiation from the inner region dominates and at longer
wavelengths, the emission arises from cold dust deeper inside the disk
and from further out.

\subsection{AGN Dust Properties \label {dust.sec}}

Dust in dense environments such as in the circumnuclear regions of AGN
is certainly different from the dust in the diffuse ISM. The variation
is caused by the huge difference in the density in both media and for
AGN some observational evidence of a change in the dust properties is
presented by \cite{M01}. The basic process for this modification of
the dust is probably grain coagulation which leads to fluffy particles
and grain growth \citep{MW89}.

The 2175\,\AA \/ extinction bump, which is attributed to small
    graphites and PAHs, is not detected in AGN \citep{M01}.  The lack
    of PAH emission in the proximity of the AGN was used as an
    indication that particles smaller than 10\,nm are destroyed
    (\citealp{Voit,Mason07}). However, \cite{AH14} and \cite{Ramos14b}
    detected in 6 Seyfert galaxies the 11.3\,$\mu$m PAH feature in the
    nuclear region.  They show that PAH molecules can survive in the
    nuclear environments of low luminosity AGN as close as 10\,pc from
    the central source.  They interpret this as evidence that the PAHs
    are not destroyed in the torus by the hard radiation of the AGN.

    Inner radii derived from K-band reverberation mapping of type 1
    AGN are a factor of 3 smaller than what is expected from standard
    ISM dust. \cite{Kishimoto07} suggest that this is due to the fact
    that the grains are larger than standard ISM dust.  However,
    \cite{Hoenig10} conclude that the grain size distribution in AGN
    is the same as that usually adopted for the ISM.  Previous models
    of AGN tori assumed diffuse ISM dust, graphite dominated dust, and
    dust with different optical constants than usually assumed for
    ISM dust.  Fritz et al. (2006) shows that variations of the
    silicate feature strength may be explained by changing the optical
    constants of the dust or its distribution in the torus. 

In this study we consider fluffy mixtures of silicate and amorphous
carbon grains \citep{KS94}. Optical constants of the materials are
from \cite{Z96} and \cite{D03}, respectively.  The bulk density of the
dust materials is 2.5\,g/cm$^{3}$, the carbon--to--silicate abundance
ratio is 6.5, and 50\% of the volume fraction of fluffy particles is
vacuum. We consider large particles with radii $a$ between 16 and
260\,nm. They follow a grain size distribution with number density
$\propto a^{-3.5}$ \citep{M77}. The influence of the dust cross
section on the AGN emission is exemplified in
Fig.~\ref{FluffyISM.pdf}. For comparison we apply dust cross sections
of pure (unmixed) particles that we call ISM dust \citep{S14}. Besides
the grain structure, ISM grains are otherwise identical to fluffy
grains. They have the same optical constants, abundances, and
sizes. Fluffy grains are more efficient absorbers and have lower
scattering cross sections than ISM dust for the whole spectrum. The
extinction cross section of fluffy grains in the V band is $K_{\rm V}
= 33.3$\,(g--dust/cm$^3$). This is a factor of 2 larger than for ISM
dust. At 1\,mm the cross section of fluffy grains is 12 times that of
ISM dust.  For the purpose of a comparison we take an AGN library
model, computed using fluffy grains, inner radius $r_{\rm {in}} =
0.25$\,pc, cloud filling factor $\eta = 1.5$\,\%, cloud optical depth
$\tau_{\rm {V, cl}} = 4.5$, and optical depth of the disk midplane of
$\tau_{\rm {V, mid}}=1000$. A model with the same dust density
distribution is run assuming ISM dust.  For the latter model the disk
midplane optical depth reduces to $\tau_{\rm {V, mid}}=750$ and the
optical depth of the clouds to $\tau_{\rm {V, cl}} = 3.375$. Fluffy
grains emit more strongly in the far IR and submillimeter than ISM
dust and the SED peak wavelength shifts towards longer wavelengths.
The cross-over wavelength where the AGN emission in face--on and
edge--on views is identical, in other words the wavelength where the
torus emission becomes isotropic, occurs for fluffy grains in the
submillimeter ($\sim 200$\,$\mu$m) and for ISM dust in the far IR
($\sim 100$\,$\mu$m).

In face--on view the emission in the mid- and near IR is similar
whereas in edge--on view the small difference in the extinction of
both dust models leads to a significant change in the continuum
flux. In the silicate band at 10\,$\mu$m the fluffy grains have a
shallower and broader absorption cross section when compared to ISM
dust. This is also reflected in the SED that is shown in
Fig.~\ref{FluffyISM.pdf}. In edge--on view the silicate absorption is
noticed using ISM dust and not when fluffy grains are considered.  In
face--on view the silicate feature has a peak emission at 10.4\,$\mu$m
for ISM dust and at 11.5\,$\mu$m for fluffy grains. In the dense dust
environment of the AGN the particles have a higher sticking
probability than in the diffuse ISM and form rather fluffy dust
agglomerates than dust in the diffuse ISM.  The 2175\,\AA \/
extinction bump is not observed in AGN spectra and therefore we use in
both dust models amorphous carbon that does not produce such a
feature. In the scattering cross section ISM dust shows a feature at
0.2\,$\mu$m \citep{H09} that is not present for fluffy grains. Such a
scattering peak is not detected for AGN.


\begin{figure*}
\centering \includegraphics[scale=0.375]{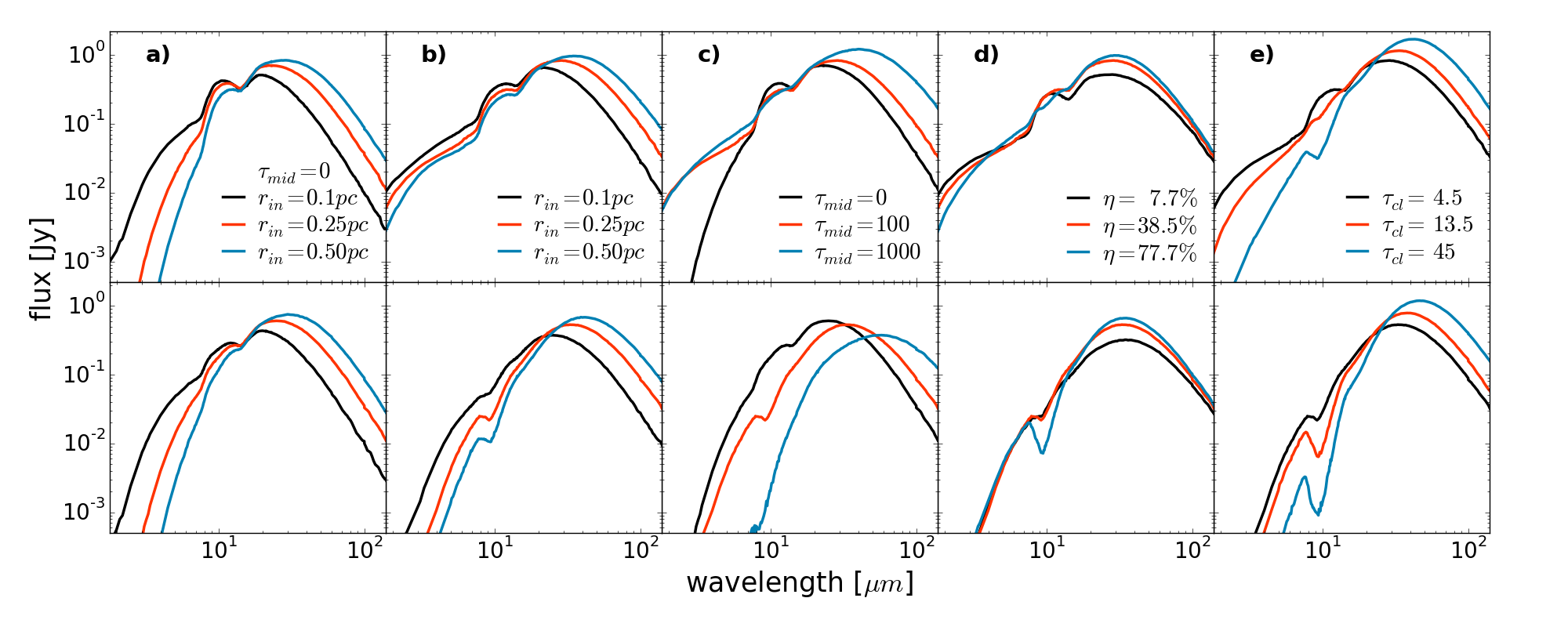}
\caption{Impact of AGN parameters on the SED in face--on (top) and
  edge--on (bottom) view.  We vary the inner radius $r_{\rm {in}}$ for
  models {\bf {a)}} without and {\bf {b)}} with a disk, {\bf {c)}} the
  optical depth of the disk midplane $\tau_{\rm {V, mid}}$, {\bf {d)}}
  the volume fraction of the clouds $\eta$, and {\bf {e)}} the optical
  depth of the clouds $\tau_{\rm {V, cl}}$. Unless otherwise specified
  $r_{\rm {in}} = 0.25$\,pc, $\tau_{\rm {V, mid}} = 100$, $\eta =
  38.5$\%, and $\tau_{\rm {V, cl}} = 4.5$. \label{all.pdf} }
\end{figure*}

\subsection{AGN Extinction \label{extinction.sec}}

The fact that dust near AGN is different from the diffuse ISM may be
best witnessed from changes in the extinction curve. However, we point
out some observational difficulties. For unresolved observations,
photon scattering within the beam may increase the detected flux
altering the wavelength dependence of the extinction, and for clumpy
media there are light paths towards the source at different optical
depths. These effects are known to change the appearance of extinction
curves and to account for this the effective optical depth is
introduced \citep{NP84, K09}:

\begin{equation}
\tau_{\rm {eff}} = - \, \ln{\frac{F_{\rm {obs}}}{F_{\rm {nd}}}}
\label{taueff.eq}
\end{equation}

where $F_{\rm {obs}}$ is the observed flux and $F_{\rm {nd}}$ is the
flux that would be observed in the absence of dust. Both quantities
are easily derived by counting the number of photons within the
aperture of an AGN model and comparing them to an identical MC run but
without dust. We illustrate this for both dust models of
Sect.~\ref{dust.sec} and take as the AGN model: $r_{\rm {in}} = 0.25$\,pc,
$\tau_{\rm {V, cl}} = 4.5$, $\tau_{\rm {V, mid}}=1000$, $\eta = 1.5$\,\%,
and in addition a disk--only model without clumps ($\eta = 0$). For
these four models it was necessary to improve the photon statistics in
the optical. We do this by launching 40 times more packets than for
the SEDs computed for the library.

\begin{figure}
\centering

\includegraphics[scale=0.69,clip=true,trim=4.6cm 9.5cm 1cm 10.cm]{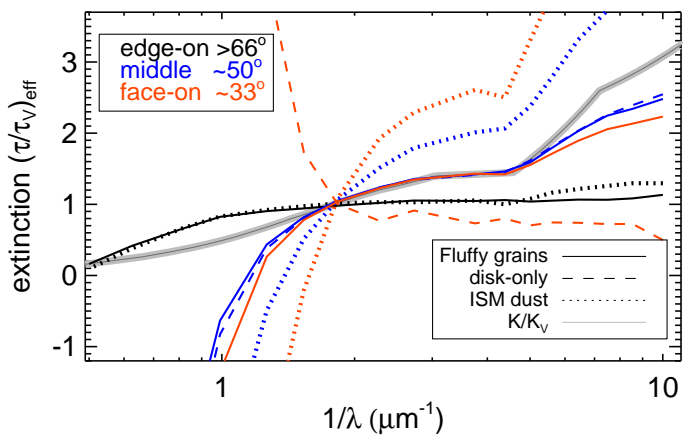}

\caption{Wavelength dependence of the effective extinction curves for
  three viewing directions: edge--on (black), intermediate (blue) and
  face-on (red) of AGN models as described in the text. The effective
  extinction curves of the AGN model computed with fluffy grains are
  shown as full line and for disk-only models as dashed lines, that of
  ISM dust as dotted lines. The extinction curve (thick gray) derived
  from the cross section of the fluffy grains is given for
  reference. All curves are normalized to the V
  band. \label{taueff.pdf} }
\end{figure}

The effective extinction curves are displayed in Fig.~\ref{taueff.pdf}
for three directions: face--on ($\theta \sim 33^{\rm o}$), an
intermediate case with $\theta \sim 50^{\rm o}$, and edge--on ($\theta
\simgreat 66^{\rm o}$).  For reference we also show the extinction
curve as derived from the dust cross section $K/K_{\rm V}$ of the
fluffy grains.  In pencil--beam observations the effective extinction
and the dust extinction are identical. It is striking that none of the
effective extinction curves resembles the one of the dust model. There
is no direct one-to-one link between the observed
extinction curve and the wavelength dependence of the dust cross
sections (Scicluna \& Siebenmorgen 2015, subm.).

In edge--on view $\tau_{\rm {V, eff}} > 0$ from the UV down to the near
IR.  In the UV the extinction curve flattens, stays below the
extinction of the dust model and becomes even gray in accordance with
the findings by \cite{NP84} and \cite{W98}. From similar observational
results it was claimed that the dust composition in the circumnuclear
region of AGNs could be dominated by large grains \citep{M01}. As
demonstrated in Fig.~\ref{taueff.pdf} such a statement does not
necessary hold. One cannot directly conclude that there is an increase
of grain size when observing a flatter than usual extinction curve and
this is especially the case when the object remains highly unresolved.

The effective extinction may also become negative.  This occurs
whenever there is more light detected with dust than without dust
\citep{K09}.  Negative extinction can be measured when extra light is
scattered into the beam and overshines the primary source. For
intermediate or face--on views presented in Fig.~\ref{taueff.pdf} this
occurs at wavelengths $\simgreat 0.8$\,$\mu$m. In this case there is
an optically thin view to the source and in addition light is
scattered from the disk into the beam.  In the face--on view of the
disk--only model the enhanced scattering contribution is important
already in the V band so that $\tau_{\rm {V, eff}} < 0$. This explains
the behaviour of the dashed red line in Fig.~\ref{taueff.pdf} and
remains stable as long as $\theta \simless 44^{\rm o}$ (close to
face--on).  For edge--on views variations in $\theta$ are unimportant
as long as the disk provides substantial extinction. For clumpy models
and viewing angles that do not hit the disk ($\theta < 60^{\rm o}$)
small variations in $\theta$ have a large impact on $\tau_{\rm {V,
    eff}}$, because by chance a clump may obscure the source or not.

The effects described above become more pronounced for ISM dust
(dotted) than for the fluffy grains because the former have stronger
scattering efficiencies (Fig.~\ref{FluffyISM.pdf}, top). They also
produce significant steeper extinction curves than is expected from
the dust model, in agreement with the qualitative estimates derived by
\cite{K09}.


\section{Impact of parameter variation on the SED}

We illustrate in Fig.~\ref{all.pdf} changes in the SED when one
parameter of the AGN library is varied while all the others stay
fixed.  For a particular model the SED in face-on view is shown in the
top and in edge--on view in the bottom panel of Fig.~\ref{all.pdf}.  In
both views, and respective panels, the SEDs are identical when the AGN
emits isotropically. This happens at wavelengths long-ward of the far
IR peak flux. The AGN torus emits an-isotropically when the SED
strongly depends on the viewing direction. This occurs at wavelengths
short-ward of the far IR peak. Panels a) and b) of Fig.~\ref{all.pdf}
illustrate how an increase of the inner radius shifts the far IR peak
to longer wavelengths. When the inner radius becomes larger the dust
gets colder, hence shows a maximum emission at longer wavelengths so
that in the far IR and submillimeter the flux increases while in the
near IR the flux decreases.

Clumpy models without and with a homogeneous disk are shown in
Fig.~\ref{all.pdf}.a,b, respectively. In face-on view models with a
disk show in the near IR a huge flux increase when compared to models
without a disk.  The addition of disk material increases the total
dust mass and this leads to an increase of the submillimeter flux at
200\,$\mu$m by a factor $2 - 3$. The impact of the midplane optical
depth of the disk $\tau_{\rm {V, mid}}$ is shown in
Fig.~\ref{all.pdf}.c. For edge--on views the disk provides additional
dust extinction changing the silicate band from emission to
absorption, while for face-on views the influence of the disk on the
strength of that feature is small. When disk material is included
there is additional hot dust in the system, and in face-on views the
near IR ($\sim 2$\,$\mu$m) flux increases by orders of magnitude
(Fig.~\ref{all.pdf}.c, top). The near IR flux may increase even
further when a puffed-up inner rim, as known to exist in
proto-planetary disks, is considered \citep{D02, SH12}. The disk
reduces the near IR flux for edge--on views provided that the
extinction is high enough ($\tau_{\rm {V, mid}} >30$,
Fig.~\ref{all.pdf}.c, bottom).  The optical depth of the clouds
$\tau_{\rm {V, cl}}$ also has a huge impact on the near IR and
submillimeter fluxes (Fig.~\ref{all.pdf}.e). In face-on view a change
of the silicate band from emission to absorption is noticed when the
optical depth of the clouds $\tau_{\rm {V, cl}}$ is increased
(Fig.~\ref{all.pdf}.e).  For edge--on views an increase of $\eta$ will
increase the dust extinction and hence produce a stronger 10\,$\mu$m
absorption feature (Fig.~\ref{all.pdf}.d, bottom). The impact of
parameter variation on the silicate band is further discussed in
Sect.~\ref{Si.sec}.

\begin{figure*} [htb]
\begin{center}
 \includegraphics[scale=1.08,clip=true,trim=0.8cm 0cm 0.8cm 15.cm]{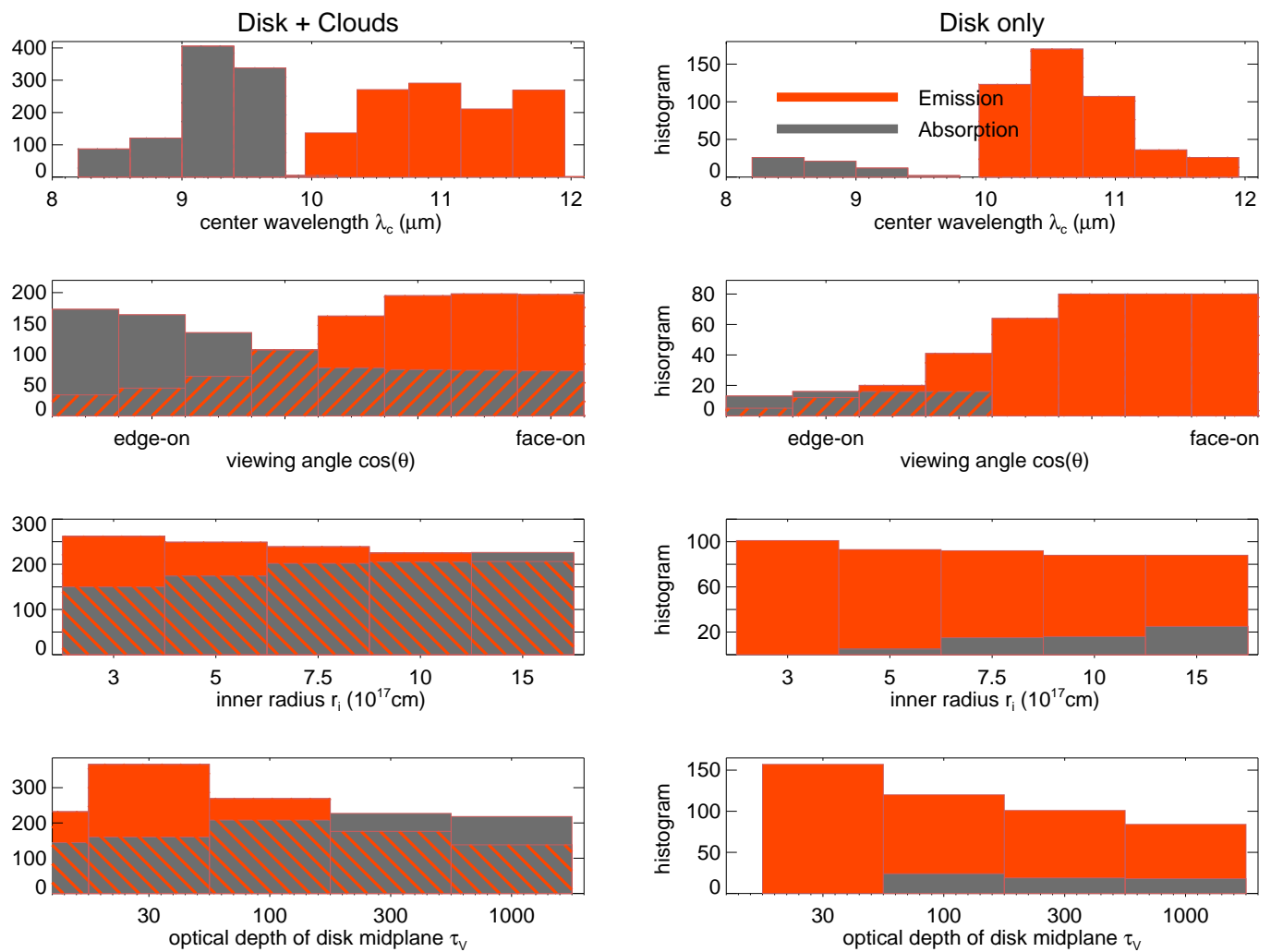}
\end{center}

\caption{Histograms of the silicate emission (red) and absorption
  (grey) band for the AGN library labeled ``Disk+Clouds'' (left) and
  disk--only models (right). Distribution functions are shown from top
  to bottom for the the central wavelength of the feature, the viewing
  angle, the inner radius of the AGN dust torus, and the optical depth
  along the midplane of the homogeneous disk \label{plSi8.pdf}. }
\end{figure*}



\subsection{Silicate Feature \label{Si.sec}}


Mid IR spectra of AGN show silicate features in absorption \citep{R91,
  Spoon02, SKS04, L07} or in emission \citep{S05, H05, H07, Sturm05,
  Sirocky08}.  \cite{Hatz15} present a census of the silicate features
in mid IR spectra of 800 AGN observed with Spitzer.  Where the feature
is in emission, in about 65\% of the cases the peak wavelength is
$>10.2$\,$\mu$m whereas the shift of the 9.7\,$\mu$m absorption
features is much smaller. The 10\,$\mu$m silicate emission and
absorption features are also observed from the ground
(\citealp{Roche06, H10,AH12,Sales11, Esquej14,
  GonzalesMartin13,Ramos14b,Ruschl-Dutra14}), and it is noticed that
the depth of the 10\,$\mu$m silicate absorption band is enhanced
compared to that measured by Spitzer.

In the MC code we applied a fine wavelength sampling
that allows us to discuss the behaviour of the 10\,$\mu$m band.  The dust
cross section of the 10\,$\mu$m silicate band is larger than that of the
18\,$\mu$m feature (Fig.~\ref{FluffyISM.pdf}).  The feature strength can
be quantified similarly to the definition of Eq.~(\ref{taueff.eq}) by
its effective optical depth

\begin{equation}
\tau_{\rm {Si}} = - \ \ln \left( \frac{F_{\rm {peak}}} {F_{\rm
    {cont}}} \right)
\label{tausi.eq}
\end{equation}

where $F_{\rm {peak}}$ is the maximum or minimum of the observed
in--band flux at center wavelength $\lambda_{\rm c}$, and $F_{\rm
  {cont}}$ is an estimate of the underlying continuum at that
wavelength. We estimate the continuum for each model spectrum by a
straight line with anchors set outside the red and blue wing of the
band at wavelengths between $5 - 7$\,$\mu$m and $12 - 14$\,$\mu$m. In
this definition, emission features appear at $\tau_{\rm {Si}} < 0$ and
absorption features at $\tau_{\rm {Si}} > 0$. Clumpy media offer some
direct views to the inner hot regions that fill the absorption and
reduce the depth of the band. The feature strength depends on the
distribution of the dust within the AGN torus \citep{L07}, the dust
mineralogy \citep{Sirocky08}, the grain structure
(Fig.~\ref{FluffyISM.pdf}), and is rather insensitive to the source
spectrum. Note that, similarly to the discussion of the extinction
curve (Sect.~\ref{extinction.sec}), the effective optical depth
derived from the 10\,$\mu$m silicate absorption feature does not
provide a measure, and at best is only a lower limit, of the optical
depth of the dust.

\begin{figure}  [htb]
 \includegraphics[scale=0.55,clip=true,trim=0.5cm 0.cm 0cm 16.5cm]{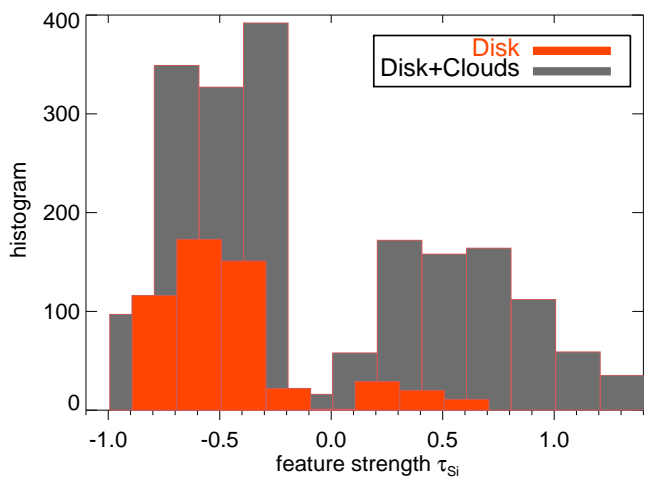}
 \caption{Distribution of the strength of the 10\,$\mu$m silicate band
   (Eq. ~\ref{tausi.eq}) for the AGN library (labeled ``Disk+Clouds'',
   gray) and for the disk--only models (red). Emission features have
   $\tau_{\rm {Si}} < 0$ and absorption features $\tau_{\rm {Si}} >
   0$.  The low frequency tail of the distribution function is
   truncated at $\tau_{\rm {Si}} = 1.4$ \label{pl_Sistrength.pdf} }

\end{figure}

Histograms of the strength of the 10\,$\mu$m silicate band are shown in
Fig.~\ref{pl_Sistrength.pdf} for the AGN library as a whole and disk--only
models. The latter are selected from the library of models without clumps
($\eta = 0$).

For Seyfert 1s the observed feature strength ranges for emission bands
between $-0.15 \leq \tau_{\rm {Si}} \leq -0.01$ and for absorption
bands between $0.06 \leq \tau_{\rm {Si}} \leq 1.5$, for quasars
between $-0.3 \leq \tau_{\rm {Si}} \leq -0.01$ and $\tau_{\rm {Si}}
\leq 0.62$, respectively \citep{Th09}. \cite{Sch08} observe a strength of
the absorption band for type 2 AGN between $0.2 \leq \tau_{\rm {Si}}
\leq 3.5$. The AGN library accounts well for the observed scatter
(Fig.~\ref{pl_Sistrength.pdf}). In the setup of disk--only models
absorption features stronger than $\tau_{\rm {Si}} \geq 0.8$ are not
present. \cite{L07} report a mean feature strength for Seyfert 1s
around $\bar{\tau}_{\rm {Si}} \sim - 0.2$ and for Seyfert 2s around
$\bar{\tau}_{\rm {Si}} \sim 0.2$. Average values derived from the
distributions functions presented in Fig.~\ref{pl_Sistrength.pdf} are
about the same. We cannot identify a striking need to postulate clumpy
AGN models when explaining observed feature strengths of the 10\,$\mu$m
silicate band. Arbitrary clump distributions ease accounting for
silicate features observed at any strength and viewing angle of the
AGN.

Disk configurations where the 10\,$\mu$m band is in absorption or
emission can be seen in the distribution functions presented in
Fig.~\ref{plSi8.pdf}. The distribution of the central wavelength
position $\lambda_{\rm c}$ of the silicate band is shown in the top
panels of Fig.~\ref{plSi8.pdf}.  In the silicate band the cross
section $K_{\rm {Si}}$ peaks at 9.5\,$\mu$m and the exact position
depends on the choice of optical constants.  Absorption features
detected in the SED match this position with some scatter that is
explained by radiative transfer effects. Noticeably for most models
$\lambda_{\rm c}$ of the emission feature is shifted to
$10.5 - 11.9\,\mu$m.  The shift of the extrema to such long
wavelengths is naturally explained assuming that the emission feature
is optically thin, so that the observed flux is estimated by
$F \sim K \times B(T_{\rm d})$.  The folding of $K$ with the steeply
rising Planck function $B(T_{\rm d})$ of the dust at temperature
$T_{\rm d}$ results in the observed wavelength shift
(\citealp{S05,Nikutta09}).

The silicate feature may appear in emission whenever there is an
optically thin view to the inner, hot regions of the disk. This may
happen for any $\tau_{\rm {V, mid}}$ of the disk midplane, inner radii
and viewing directions (Fig.~\ref{plSi8.pdf}, right).  Most of the
emission bands in disk--only models are seen close to face--on
views. There are only a few disk--only models that have sufficient
dust extinction so that the silicate band is in absorption. These are
models with $r_{\rm {in}} \geq {\rm {pc}}/4$, $\tau_{\rm {V, mid}}
\simgreat 100$, and the disk view needs to be $\theta \simgreat
60^{\rm o}$. When clumps are considered (Fig.~\ref{plSi8.pdf}, left)
the silicate band may appear either in absorption or emission for an
almost arbitrary choice of disk configurations and viewing angles. It
is striking that there are clumpy models displaying the emission band
in edge--on view. There is a rather constant decrease of the number of
'Disk+Clouds' models that show the emission band when the inner radius
or the optical depth is increased or vice versa for the absorption
band. All this is consistent and explained by the corresponding
increase or decrease of the dust extinction and temperature.

\subsection{Apparent versus intrinsic AGN luminosity}

Because of the assumed spectral shape of the AGN (Eq.~\ref{agn.eq}),
photons are predominantly emitted from the central source at far UV to
near IR wavelengths.  Dust near the AGN scatters and absorbs these
photons and re--emits this radiation in the infrared.  By integrating
over the observed fluxes one derives the 'apparent' AGN luminosity
$L_{\rm {obs}}$. One is however interested in deriving from
observations the intrinsic luminosity of the AGN source $L_{\rm
  {AGN}}$. The AGN library provides a utility that allows the
conversion of the apparent (observed) luminosity to the {\it
  {intrinsic}} luminosity of the AGN.

\begin{figure*} [htb]
\includegraphics[scale=1.05,clip=true,trim=0.5cm 0cm 3cm 15cm]{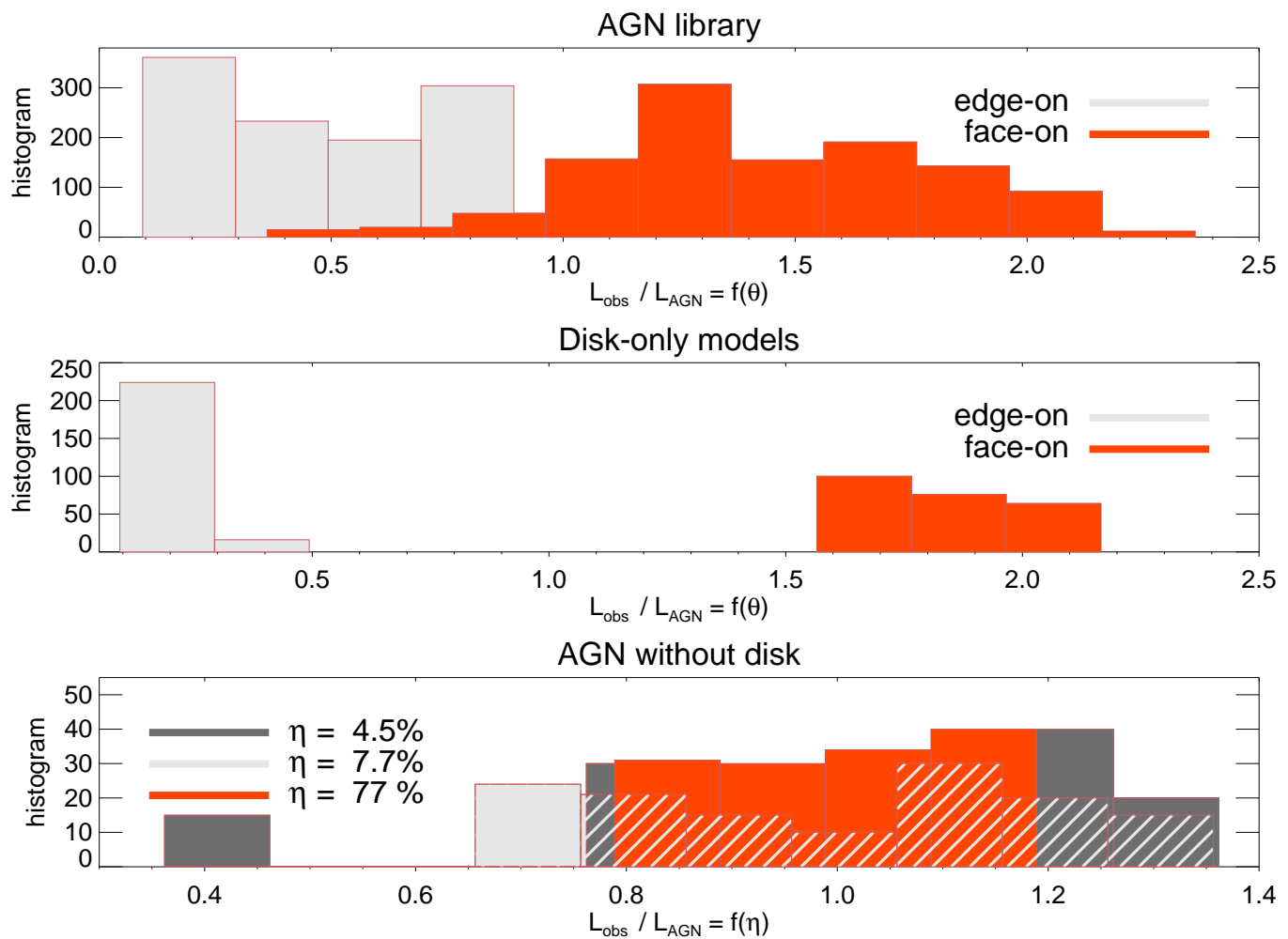}
\caption{Histograms of the apparent--to--intrinsic AGN luminosity
  (Eq.~\ref{Ldust.eq}). Results are given for the complete AGN library
  (top), and for disk-only models (middle) in face-on ($\theta
  \simless 43 ^{\rm o}$) and edge--on ($\theta \simgreat 73 ^{\rm o}$)
  views. The luminosity ratio is shown for the volume filling factor
  of the clouds $\eta$ (bottom) for models without the homogeneous
  disk. \label{Ldust.pdf} }
\end{figure*}

In the models the apparent luminosity is computed from photons emitted
into a given viewing bin.  We compute the observed luminosity for
models where dust is considered or neglected. For the latter, the
no-dust models, the luminosity is constant because the central source
emits isotropically. Therefore in one of the nine viewing bins used,
the luminosity $L_{\rm {11}} / 9 = 4 \, \pi \, D^2 \
F_{\rm{nd}}(\theta)$, where in the library $D=50$\,Mpc, and
$F_{\rm{nd}}(\theta)$ is the flux integrated over frequency in a
particular $\theta$ direction for the ``no dust'' model. For models
with dust the luminosity is computed from the flux as given in the
library, and $L_{\rm {obs}}(\theta) = 4 \, \pi \, D^2 \ F(\theta)$. We
define the anisotropy conversion factor

\begin {equation}
\epsilon = { F  (\theta) \over {F_{\rm {nd} }}} 
\label{conv.eq}
\end{equation}

The conversion factor becomes larger than 1 whenever there are more
photons detected in a viewing bin than originally emitted in that
direction from the central source. The conversion factor $\epsilon$
allows one to derive the {\it{intrinsic}} AGN luminosity $L_{\rm
  {AGN}}$. The ratio of the observed--to--intrinsic luminosity is
given by

\begin {equation}
L_{\rm {AGN} } = 9 \ {L_{\rm {obs}} (\theta)  \over \epsilon } 
\label{Ldust.eq}
\end{equation}

The distribution functions of $L_{\rm {obs} }/ L_{\rm {AGN}}$ are
shown in Fig.~\ref{Ldust.pdf} for different AGN models. Typically the
apparent luminosity is about two times larger for type Is (face--on)
than for type IIs (edge--on). The fact that for the same intrinsic AGN
luminosity, models which correspond to type 1 sources emit more IR
radiation than type 2s is clearly visible in the SEDs
(Fig.~\ref{all.pdf}). Such an increase of the IR radiation is observed
in a sample of 3CR sources \citep{S04}. For face--on views there are
often more photons radiated into the beam than originally emitted from
the source, hence $L_{\rm {obs} }/ L_{\rm {AGN}} > 1$.  Models with
clumps show for edge--on views a broader distribution in
$L_{\rm {obs} }/ L_{\rm {AGN}}$ than models without clumps. The impact
of the cloud filling factor $\eta$ is shown for models without a disk
in Fig.~\ref{Ldust.pdf} (bottom). For viewing directions with a single
cloud along the line of sight typically 40\% of the AGN luminosity is
observed. This may happen for AGN without a disk and with a small
number of clouds ($\eta = 1.5$\,\%).  When there are more clouds
$L_{\rm {obs} }/ L_{\rm {AGN}}$ increases and ranges between
$0.7 - 1.3$.

We note that the 'apparent' AGN luminosity, which is usually what is
quoted in the literature, must simply be divided by $\epsilon$ in
order to estimate the intrinsic luminosity of the AGN
Eq.~(\ref{Ldust.eq}). The conversion factor $\epsilon$ is the inverse
of the anisotropy factor $A$ defined by \cite{E06}.


\begin{figure*}
\centering
\includegraphics[scale=1.,clip=true,trim=3cm 6cm 3cm 6cm]{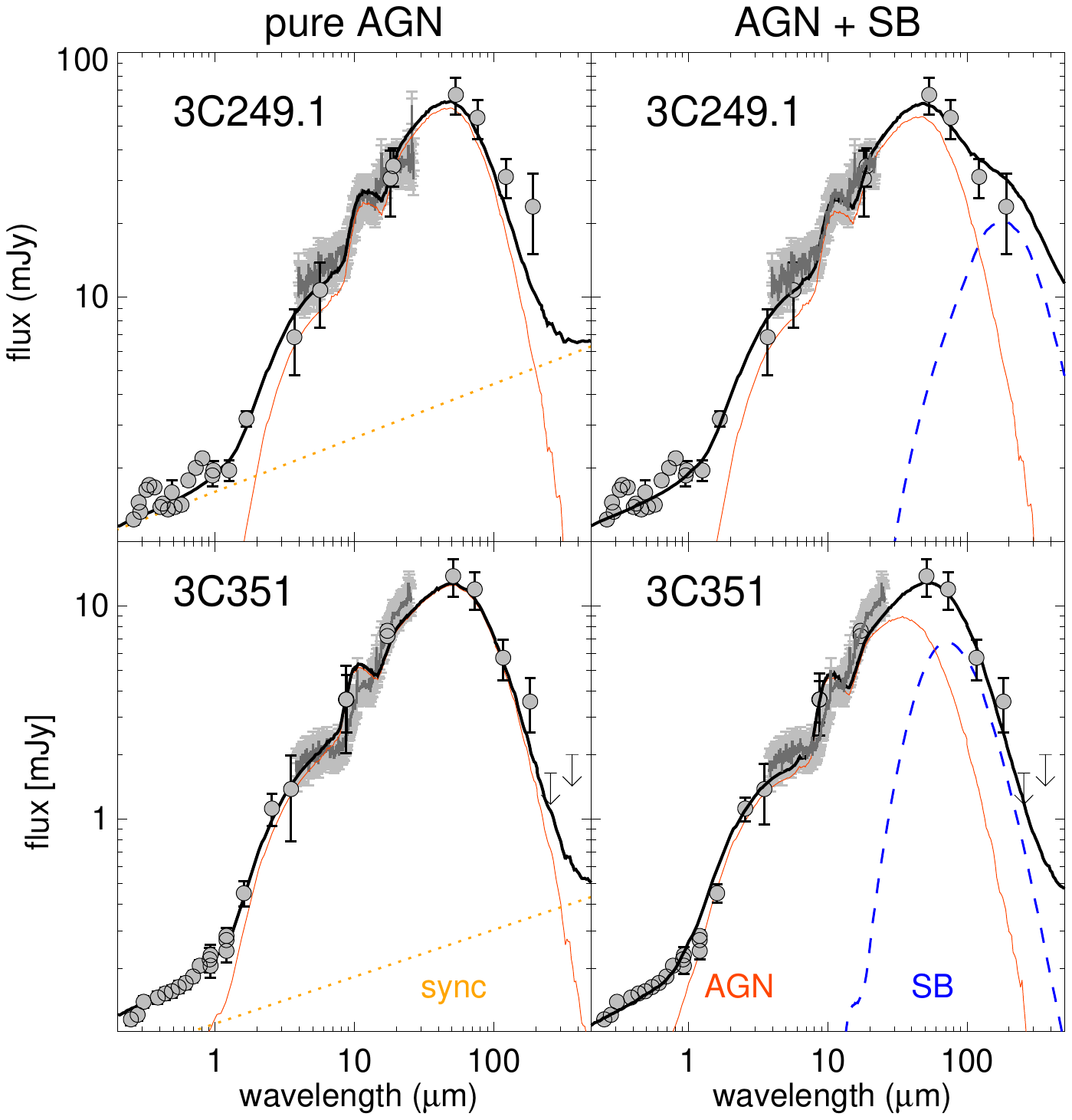}
\caption{Rest frame SED of the quasars 3C249.1 (top) and 3C351
  (bottom). Data are from the NASA/IPAC Extragalactic Database (NED),
  Spitzer/IRS \citep{L10} and Herschel ({\it {this work}},
  Table~\ref{photo.tab}).  A fit using a pure AGN is shown on the left
  and for a combination (black line) of AGN (red line) and starburst
  (blue dashed) activity on the right.  A synchrotron component
  (yellow dots) is added to better match the near IR. Model parameters
  are given in Table~\ref{para.tab} \label{allQuasar_sed.pdf}.}
\end{figure*}

\begin{table}
\begin{center}
  \caption {Herschel photometry with 1$\sigma$ flux
    uncertainty.  \label{photo.tab}}
 \begin{tabular}{|l|r|r|}
   \hline
   & &   \\
   Object & Filter & Flux \\
          &  ($\mu$m) & (Jy)   \\
\hline
NGC~1365& 70 &$128 \pm 14$ \\
  & 100  &$148  \pm 16$ \\
  & 160  &$179  \pm 20$ \\
\hline
NGC~3081& 70 &$2.8 \pm 0.3$ \\
  & 100  &$3.1 \pm 0.4$ \\
  & 160  &$3.5 \pm 0.4$ \\
\hline
NGC~4151& 70 &$5.2 \pm 0.6$ \\
  & 100  &$3.4 \pm 0.4$ \\
  & 160  &$4.9 \pm 0.6$ \\
\hline
   & &   \\
          &  ($\mu$m) & (mJy)   \\
   \hline
3C249.1 & 70 &$87.9 \pm 6.2$ \\
 & 100 &$70.9 \pm 5.7$ \\
 &160  &$40.6 \pm 3.3$ \\
 &250  &$30.7 \pm 8.0$ \\
 &350  &$41.7 \pm 8.7$ \\
 &500  &$50.7 \pm 9.4$ \\
   \hline
3C351& 70 &$189.6 \pm 19.1$ \\
  & 100  &$164.4 \pm 16.2$ \\
  & 160  &$78.5 \pm 9.3$ \\
 & 250 &$48.9 \pm 9.1$ \\
 & 350 &$19.6 \pm 7.5$ \\
 & 500 &$16.5 \pm 9.2$ \\
\hline
\hline
\end{tabular}
\end{center}
\end{table}

\section{Testing the SED library}

In this section we fit the observed SEDs of a number of representative
objects from the near IR to the submillimeter including Spitzer/IRS
spectroscopy, Herschel photometry of distant high luminosity AGN and
high spatial resolution NIR and MIR observations of nearby, low
luminosity Seyfert nuclei.  Further we test that sources of the
library would be classified as AGN when applying the empirical mid IR
selection criteria by \cite{Stern05}.  First, we test how much of the
observed SED can be explained by pure AGN activity \citep{Fritz06}. As
a second step, we combine the predicted SED from the AGN library with
an additional starburst (SB) component.  We call such combination
'AGN+SB' models. To this end we select elements from the SED library
of starburst models of \cite{SK07}. We consider the starbursts radius
$R_{\rm {SB}}$, the luminosity $L_{\rm {SB}}$, the visual extinction
$A_{\rm {V}}$, and the ratio of the luminosity that is due to OB stars
$\eta_{\rm {OB}}$ as free parameter and keep the hot spot density
constant $\eta_{\rm {HS}} = 10^4$\,cm$^{-3}$. In addition, to better
match data below 2\,$\mu$m we add an extra component. For quasars we
use beamed synchrotron emission, which is approximated by a power-law,
and for the galaxies stellar light from the host, which is
approximated by a blackbody. The extra components add a minor
contribution to the photometry between the thermal IR and the
submillimeter. Nevertheless we subtract the flux of the extra
components from the data when fitting the dust emission. For the
latter fit we use a Bayesian formalism that utilizes the Markov Chain
Monte Carlo algorithm by \cite{Johnson13}. This general purpose SED
fitting tool employs the Metropolis–Hastings algorithm,
e.g. \cite{Metropolis}. The code is able to take any set of SED
templates. In our case we use either the pure AGN library or
simultaneously the combination of the AGN+SB libraries. The tool
returns the requested best-fitting parameter estimates, and in
addition construct the confidence levels from the posterior parameter
distribution. We verified the accuracy and robustness of that SED
fitter against results obtained with traditional least--squares
methods. We confirm \cite{Johnson13} that the programme recovers
best-fitting values similar to those from traditional methods.

\begin{figure*}
\centering
\includegraphics[scale=1.,clip=true,trim=3cm 6cm 3cm 6cm]{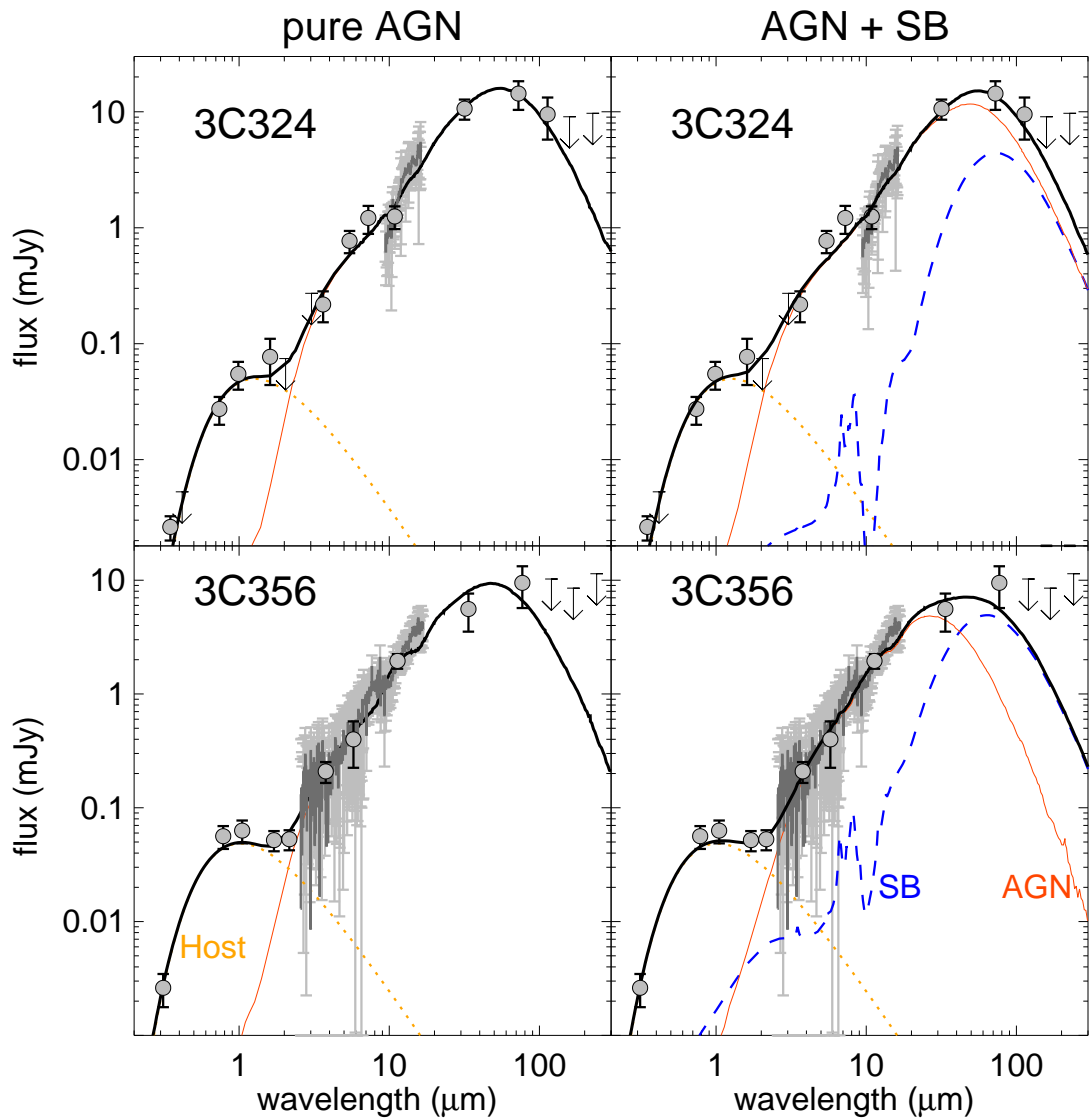}
\caption{Rest frame SED of the radio galaxies 3C324 (top) and 3C356
  (bottom). Data are from NED, Spitzer/IRS \citep{L10} and Herschel
  \citep{Podi15}. A fit using a pure AGN is shown on the left and for
  a combination (black line) of AGN (red line) and starburst (blue
  dashed) activity on the right. Stellar light from the host (yellow
  dots) is added to better match the near IR. Model parameters are
  given in Table~\ref{para.tab}. \label{allRG_sed.pdf}}
\end{figure*}

\begin{figure*}
\centering
\includegraphics[scale=1.2,clip=true,trim=3.4cm 9.5cm 3.4cm 10cm]{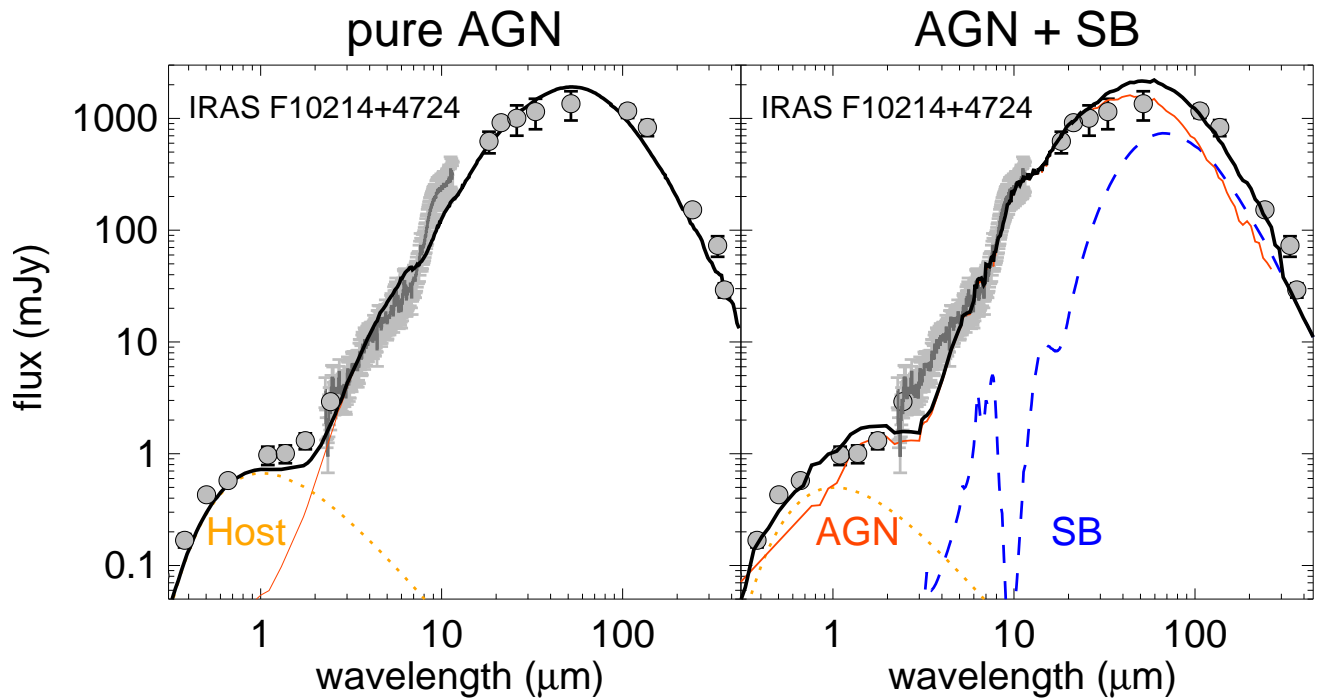}
\caption{As Fig.~\ref{allRG_sed.pdf} for the hyperluminous infrared galaxy
  IRAS10214+4724 with data from NED, Spitzer IRS \citep{HC11}, and
  MIPS, IRAC and Herschel by \cite{E13}. A fit using pure AGN models
  are shown with (dashed) and without (dotted) clumps. Model
  parameters are given in
  Table~\ref{para.tab}.  \label{allIras_sed.pdf}}
\end{figure*}

In the fit procedure the Spitzer spectroscopy data are binned
to a spectral resolution that adequately reproduces the shape of the
SED including the silicate feature.  We display the SED in
Figs.~\ref{allQuasar_sed.pdf} -- \ref{allIras_sed.pdf} and the best
fit parameters including their uncertainties are summarised for the
pure AGN models in Table~\ref{para.tab} and for the combination of AGN
and starburst activity in Table~\ref{paragnsb.tab}. Apart from the
inclination, the fit parameters may cover almost the entire parameter
range because of the arbitrary cloud distribution.  One finds adequate
fits (solutions) of the SED using either a combination of a large
number of clouds with small optical depth, or models that have a small
number of clouds with high optical depth. Such a change of the cloud
distribution requires that the inner radius, the optical depth of the
disk or other of the free parameters are modified as well.

\subsection{Observations}

We derive aperture photometry on the final images extracted from the
Herschel archive and apply correction factors for the encircled energy
fraction and colors as given by \cite{Balog14} and the User
Manuals. The background is subtracted and estimated from the mean flux
that is measured in a 2$''$ wide annulus which is placed 1$''$ further
out than the source aperture.

PACS and SPIRE photometry of 3C249.1 and 3C351 (PI: Luis Ho) is
available in 6 Herschel filters and summarised in
Table~\ref{photo.tab}. The total flux is measured and uncertainties
are computed by adding on the statistical error a calibration
uncertainty of 7\% for PACS and 5.5\% for SPIRE bands as well as a
confusion noise of 5.8, 6.3, and 6.8\, mJy in the bands at 250, 350,
and 500\,$\mu$m, respectively.

PACS photometry of the Seyfert nuclei NGC~1365 {\bf {(PI: M. Sanchez)}},
NGC~3081 {\bf {(PI: M. Sanchez)}}, and NGC~4151 {\bf {(PI: A. Alonsoh)}}
is measured by adjusting a radius of the source aperture that matches
the diffraction limit of the Herschel telescope.  We apply source
radii of $6''$, $7''$, and $12''$ at 70, 100, and 160\,$\mu$m,
respectively \citep{Balog14}. We centre this aperture on the location
of the AGN.  These galaxies are resolved by Herschel but their AGN
tori are not. For the other objects Herschel photometry is obtained
from \cite{E13} and \cite{Podi15}.


\subsection{Fits to the quasars 3C351 and 3C249.1}


The near IR photometry of 3C351 is fit assuming a contribution of a
beamed synchrotron component that we approximate as a power-law
($F_\nu \propto \nu^\alpha)$ with a slope $\alpha \sim -0.2$. We show
that pure AGN models with parameters are summarized in
Table~\ref{para.tab} account for the complete SED without the need of
postulating additional starburst activity.

The best fit of the AGN+SB models shows that the SED can be explained
mainly by a torus and a small (8\,\%, Table~\ref{paragnsb.tab})
contribution from a starburst.  The predicted viewing direction is
close to face on ($\theta \sim 33^{\rm o}$).  The models reproduce the
silicate emission feature of this quasar and the near IR bump without
the need for an additional hot dust component as suggested by
\cite{Mor09}. This is a direct consequence of additional material in
the innermost region of the torus that is provided by the homogeneous
disk and by clouds in the ionisation cones of the AGN.  We note that
our conclusion that the starburst makes a negligible contribution to
the bolometric luminosity is a consequence of the large ratio of
$r_{\rm {out}}/r_{\rm {in}} \sim 170$, and the assumed fluffy grain
model.  The far IR emissivity of fluffy grains is much larger than
that of ISM dust (Sect.~\ref{dust.sec}, Fig.~\ref{FluffyISM.pdf}).

For 3C249.1 we find that pure AGN models provide a reasonable fit to
the SED and its silicate emission feature without the need for SB
activity. There are of course AGN+SB models, similar to 3C351, that
fit the data.  Again, we assume synchrotron emission that makes a
significant contribution to the near IR. We find that in this quasar
the contribution to the total luminosity from a starburst is 3\,\%
(Table~\ref{para.tab}).

\subsection{Fits to the radio galaxies 3C324 and 3C356}

The first thing that is striking about the two radio galaxies we
studied here (3C324 and 3C356) is that their SEDs are distinctly
different from those of the quasars. They show a much weaker near- to
mid IR continuum relative to the far IR, a completely different slope
in the $1 - 10\,\mu$m range and evidence of shallow absorption
features at 10\,$\mu$m.  All of these differences are understood in
terms of the optically thick torus model and are strongly supportive
of the idea of unification. For the radio galaxies we add to the pure
AGN torus or the AGN+SB models contributions from black bodies at
temperature of $4000 - 5000$\,K, which model the emission from old
stars in the host galaxies in these objects.  The SED of both galaxies
are fit by pure AGN models with parameters as given in
Table~\ref{para.tab}. In the combined models the starburst component
is 8\,\% for 3C324 and 19\,\% for 3C356 of the intrinsic AGN luminosity
(Table~\ref{paragnsb.tab}).

\subsection{Fit to the hyperluminous infrared galaxy IRAS F10214+4724}

We discuss a fit of the SED of the hyperluminous infrared galaxy IRAS
F10214+4724 which has an apparent luminosity that exceeds
$10^{14}$\Lsun \/ \citep{RR91}. It is generally accepted that the
enormous luminosity of this galaxy at redshift z=2.224 is due to
gravitational lensing by a foreground galaxy with magnification
between 50 and 100 (\citealp{Broadhurst,Serjeant}) or $15 - 20$
\citep{Deane}. Following the detection of a silicate emission feature
\citep{T06} in this narrow-lined object, \cite{E06} and \cite{E13}
proposed that the emission feature is due to emission from narrow-line
region clouds at a temperature of 210\,K.

Here we find that the SED of IRAS F10214+4724 can also be fit by a
pure AGN model without clouds when a small inner radius or in a
'Disk+Clouds' configuration when larger inner radii are assumed
(Table~\ref{para.tab}). In the AGN+SB models we find the ratio of
starburst to AGN luminosity is 20\,\% (Table~\ref{paragnsb.tab}). The
model presented here for IRAS F10214+4724 is similar to that presented
in \cite{E13} as in both models the mid IR emission is predicted to
come from the conical region that is associated with the narrow-line
region.  The dominance of the mid IR emission of this conical region
over the emission from the main body of the disk when the system is
viewed approximately edge--on is illustrated in the image (top middle)
plotted in Fig.~\ref{images.pdf}. We also note that \cite{X14}
recently reported three other local ULIRGS with Spitzer spectroscopy
which show similar SEDs to IRAS F10214+4724.


\begin{table*}[h!tb]
\begin{center}
  \caption {Fitting parameters of pure AGN models for representative objects. \label{para.tab}}
 \begin{tabular}{|l|c||r|r|r|r|c||c|c|c|}
\hline
 \multicolumn{2}{|c||} { } &   \multicolumn{5}{|c||} { }  &   \multicolumn{3}{|c|} { } \\
 \multicolumn{2}{|c||} { } &   \multicolumn{5}{|c||} {Model Parameter}  &   \multicolumn{3}{|c|} {Derived Quantities} \\
 \multicolumn{2}{|c||} { } &   \multicolumn{5}{|c||} { }  &   \multicolumn{3}{|c|} { } \\
\hline
   & & & & & & & & & \\
Object   & Type  &Radius \/ \/ & Filling \/ & Cloud  & Disk  & Inc. &Conv.$^{\/a)}$  &Radius & Intrinsic AGN \\
& & $r_{\rm {in}}$   & factor $\eta$  &$\tau_{\rm V, cl}$ \/ \/ & $\tau_{\rm {V, mid}}$ \/ & $\theta$ & $\epsilon$& $R_{\rm {in}}^{ \/ b)}$ & luminosity L$_{\rm {AGN}}$\\
   &  & ($10^{17}$cm) &   & & &  ($^o$)& &(pc)  & log(\Lsun \/)\\
   \hline
  & & & & & & & & &  \\
   NGC~1365 & I  &7.8$^{+7}_{-5}$  & 8$^{+67}_{-7}$ & 14$^{+31}_{-9}$&470$^{+530}_{-430}$ & 33$^{+10}_{-14}$ &  1.34$^{+0.12}_{-0.02}$ & 0.1$^{+0.09}_{-0.06}$ & 10.23$^{+0.05}_{-0.04}$\\
   & & & & & & & & & \\
\hline
   & & & & & & & & & \\
   NGC~4151 & I  &4.8$^{+4}_{-2}$  & 3$^{+75}_{-2}$ & 40$^{+5}_{-39}$&230$^{+770}_{-190}$ &  43$^{+9}_{-24}$ &  2.08$^{+0.12}_{-0.02}$ & 0.1$^{+0.41}_{-0.02}$ & 10.35$^{+0.02}_{-0.02}$\\
   & & & & & & & & & \\
   \hline
    & & & & & & & & & \\
   NGC~3081 & II &3.4$^{+12}_{-3}$  & 16$^{+62}_{-14}$ & 45$^{+1}_{-44}$&7$^{+990}_{-6}$  &  80$^{+6}_{-27}$ &  0.72$^{+0.09}_{-0.02}$ & 0.1$^{+0.41}_{-0.02}$ & 11.04$^{+0.05}_{-0.05}$\\
    & & & & & & & & & \\
   \hline
    & & & & & & & & & \\
    NGC~5643 & II    & 15$^{+0.5}_{-12}$ & 48$^{+23}_{-44}$&14$^{+31}_{-10}$ &  400$^{+600}_{-400}$ & 67$^{+19}_{-7}$ &  0.84$^{+0.02}_{-0.19}$ & 0.3$^{+0.12}_{-0.2}$ & 10.65$^{+0.11}_{-0.02}$\\
    & & & & & & & & & \\
   \hline
    & & & & & & & & & \\
   3C249.1 &I & 7.7$^{+5}_{-2}$ & 2$^{+33}_{-1}$&1$^{+4}_{-1}$ &  340$^{+560}_{-240}$ & 60$^{+7}_{-7}$ &  0.82$^{+0.23}_{-0.26}$ & 4.8$^{+3.1}_{-1.3}$ & 13.57$^{+0.11}_{-0.17}$\\
    & & & & & & & & & \\
   \hline
    & & & & & & & & & \\
   3C351 & I& 7.8$^{+2.2}_{-4.5}$ & 14$^{+60}_{-13}$&1$^{+4}_{-1}$ &  840$^{+160}_{-530}$ & 43$^{+17}_{-24}$ &  2$^{+0.04}_{-0.05}$ & 4.7$^{+1.3}_{-2.7}$ & 13.54$^{+0.02}_{-0.02}$\\
    & & & & & & & & & \\
   \hline
    & & & & & & & & & \\
    3C324 & II& 15$^{+0.4}_{-12}$ & 2$^{+74}_{-1}$&40$^{+5}_{-40}$ &  290$^{+700}_{-170}$ & 67$^{+6}_{-7}$ &  0.6$^{+0.12}_{-0.14}$ & 20.3$^{+0.6}_{-16.3}$ & 14.24$^{+0.12}_{-0.08}$\\
    & & & & & & & & & \\
   \hline
    & & & & & & & & & \\
   3C356 & II & 10$^{+5}_{-7}$ & 7$^{+70}_{-5}$&4$^{+40}_{-4}$ &  260$^{+740}_{-260}$ & 67$^{+21}_{-7}$ &  0.66$^{+0.16}_{-0.18}$ & 9.2$^{+4.6}_{-6.4}$ & 13.9$^{+0.09}_{-0.14}$\\
    & & & & & & & & & \\
   \hline
    & & & & & & & & & \\
   F10214$^{\/ c)}$  & HL & 15$^{+0.4}_{-5}$ & 40$^{+37}_{-33}$&6$^{+2}_{-6}$ & 910$^{+90}_{-600}$ & 67$^{+5}_{-7}$ & 0.66$^{+0.14}_{-0.15}$ &-& -\\
   +4724 &  & & & & & & & & \\
   \hline
   \hline
\end{tabular}
\end{center} {\bf {Notes. $^{a)}$}} Anisotropic luminosity conversion
factor (Eq.\ref{conv.eq}).  $^{b)}$ Inner radius of the dust torus 
 $R_{\rm {in}} = r_{\rm {in}} \, \sqrt{L_{\rm
    AGN}/10^{11}}$. $^{c)}$ We do not estimate $R_{\rm {in}}$ and
$L_{\rm {AGN}}$  (Eq.~\ref{Ldust.eq}) for the lensed galaxy.
\end{table*}

\begin{table*}[h!tb]
\small
\begin{center}
  \caption {Fitting parameters of a combination of AGN and starburst
    models for representative objects. \label{paragnsb.tab} }
 \begin{tabular}{|l|r|r|r|r|c|c|c|c|c||c|c|c|c|}
\hline
  &   \multicolumn{5}{|c|} { }  &   \multicolumn{4}{|c||} { }  &   \multicolumn{4}{|c|} { } \\
  &   \multicolumn{5}{|c|} {AGN parameter}   &   \multicolumn{4}{|c||} {Starburst parameter}  &   \multicolumn{4}{|c|} {Derived Quantities} \\
 &   \multicolumn{5}{|c|} { }  &   \multicolumn{4}{|c||} { } &   \multicolumn{4}{|c|} { } \\
\hline
Object & $r_{\rm {in}}$ & $\eta$  &$\tau_{\rm V, cl}$ & $\tau_{\rm {V, mid}}$ \/ & $\theta$ & $R_{\rm SB}$ & $L_{\rm SB}$ & $A_{\rm V}$ & $\eta_{\rm{OB}}$ &
$\epsilon$& ${L_{\rm {SB}}}\over {L_{\rm {obs}}}^{ \/ a)}$&  $R_{\rm {in}}$ & $L_{\rm {AGN}}$\\
  & ($10^{17}$cm) &   & & &  ($^o$)& (kpc) & log(\Lsun \/) &(mag.) &(\%)& (\%) &  (\%) &(pc) & log(\Lsun \/)\\
   \hline
   & &  & & & & & & & & & & & \\
3C249.1  & 8.7$^{+0.2}_{-0.9}$ & 2$^{+21}_{-0.5}$&2$^{+0.4}_{-0.3}$ &  240$^{+310}_{-70}$ & 63$^{+2}_{-3}$ & 2.0$^{+1.0}_{-1.3}$ & 10.75$^{+0.1}_{-0.1}$ &126$^{+18}_{-97}$& 83$^{+7}_{-31}$ &  0.82$^{+0.04}_{-0.01}$ & 3.4$^{+1.7}_{-1.5}$ & 3.9$^{+1.7}_{-1.3}$ & 13.28$^{+0.02}_{-0.01}$\\
   & &  & & & & & & & & & & & \\
   \hline
   & &  & & & & & & & & & & & \\
   3C351 & 6.4$^{+0.05}_{-1.2}$ & 12$^{+11}_{-10}$&2$^{+0.4}_{-0.4}$ &  210$^{+790}_{-37}$ & 64$^{+2}_{-30}$ & 1.0$^{+1.0}_{-0.3}$ & 11.69$^{+0.1}_{-0.1}$ &83$^{+61}_{-29}$& 52$^{+8}_{-12}$&  1.51$^{+0.04}_{-0.05}$ & 5.5$^{+2.2}_{-2.4}$ & 4.9$^{+2.1}_{-1.1}$ & 13.75$^{+0.01}_{-0.01}$\\
   & &  & & & & & & & & & & & \\
   \hline
   & &  & & & & & & & & & & & \\
   3C324  & 12.1$^{+0.6}_{-4.4}$ & 2$^{+2}_{-1}$&36$^{+9.4}_{-16.8}$ &  314$^{+233}_{-141}$ & 68$^{+4}_{-3}$ & 1.7$^{+1.3}_{-1.4}$ & 11.88$^{+0.3}_{-0.9}$ &67$^{+77}_{-13}$& 46$^{+44}_{-6}$&  0.62$^{+0.02}_{-0.19}$ & 13.8$^{+3.3}_{-6.3}$ & 9.5$^{+4.4}_{-0.5}$ & 13.76$^{+0.01}_{-0.16}$\\
   & &  & & & & & & & & & & & \\
   \hline
   & &  & & & & & & & & & & & \\
3C356   & 3.2$^{+0.9}_{-0.2}$ & 1$^{+0.8}_{-0.7}$&2$^{+2.2}_{-1.9}$ &  84$^{+464}_{-18}$ & 70$^{+7}_{-4}$ & 1.5$^{+1.5}_{-1.1 }$ & 11.88$^{+0.1}_{-1.4}$ &37$^{+107}_{-28}$& 50$^{+40}_{-10}$&  0.58$^{+0.09}_{-0.06}$ & 33.3$^{+11.5}_{-17.7}$ & 1.7$^{+1.2}_{-0.4}$ & 13.45$^{+0.06}_{-0.05}$\\
    & &  & & & & & & & & & & & \\
   \hline
   & &  & & & & & & & & & & & \\
F10214    & 3.6$^{+0.5}_{-0.6}$ & 40$^{+18}_{-1.5}$&36$^{+18.1}_{-1.5}$ &  34$^{+222}_{-4}$ & 27$^{+11}_{-2}$& 1.7$^{+0.3}_{-1.0}$ & 11.88$^{+0.1}_{-0.1}$ &92$^{+51}_{-39}$& 45$^{+44}_{-5}$& 1.26$^{+0.05}_{-0.03}$ &19.1$^{+7.6}_{-8.5}$ &-&-\\
  +4724   & &  & & & & & & & & & & & \\
   \hline
   \hline
\end{tabular}
\end{center} {{\bf {Notes.}} Notation as of Table~\ref{para.tab}.$^{ \/ a)}$ Ratio of the starburst and the apparent AGN luminosity. We
    consider for the starburst activity the library by \cite{SK07}.}
\end{table*}


\begin{figure}
\centering
\includegraphics[scale=0.87,clip=true,trim=5.3cm 4cm 3cm 4.5cm]{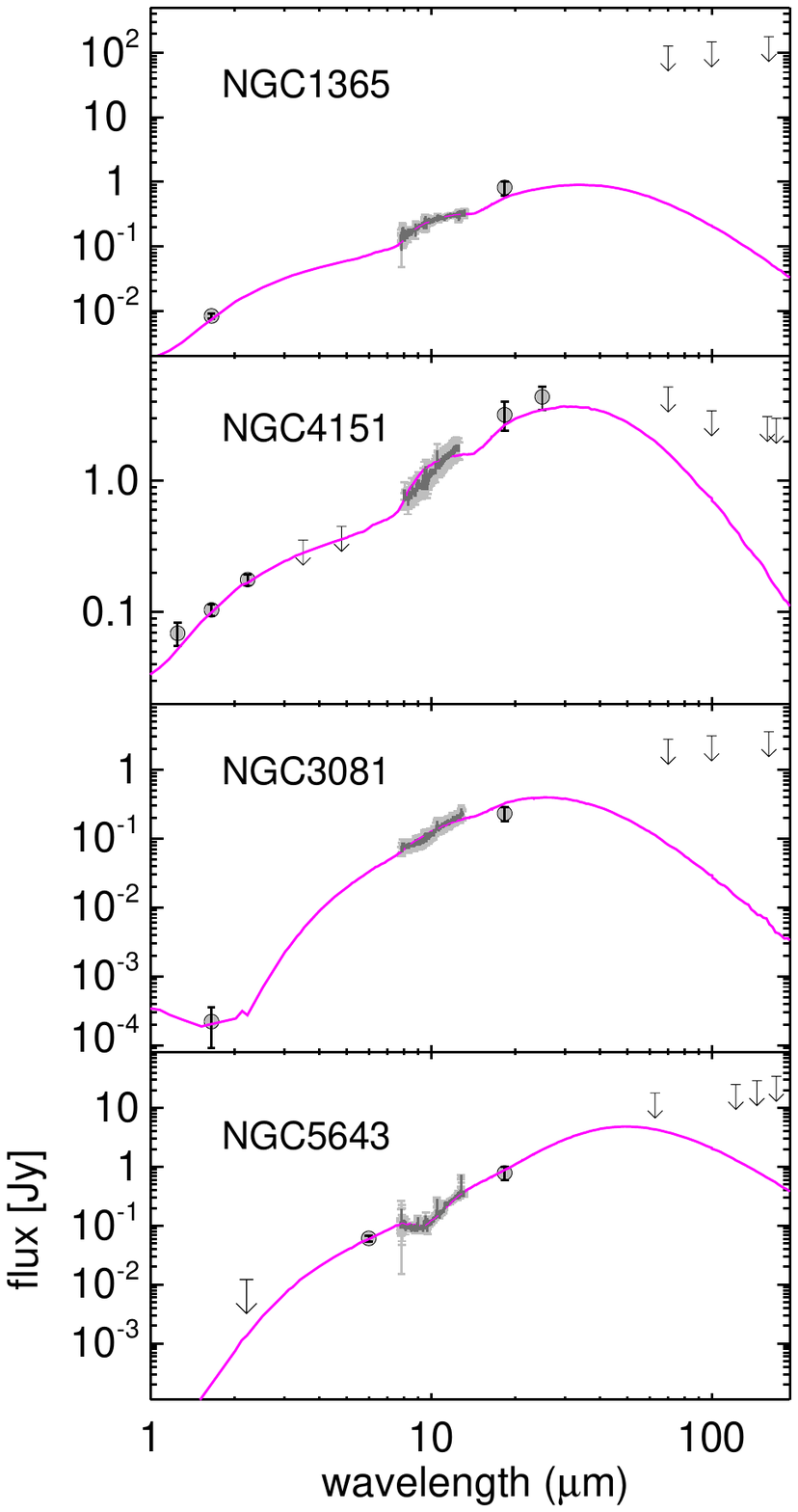}
\caption{SED fits using pure AGN models (magenta) of Seyfert nuclei:
  NGC~1365, NGC~4151, NGC~3081, and NGC~5643. High spatial resolution
  photometry is from \cite{Ramos14b}. Ground based MIR spectroscopy
  of NGC~1365 is performed by \cite{Burtscher13}, NGC~3081 and
  NGC~4151 by \cite{Ramos14b}, and NGC~5643 by \cite{H10},
  respectively. FIR photometry of the nuclear region provided by
  NED or Herschel (Table~\ref{photo.tab}) are plotted as upper limits
  of the AGN torus emission. Model parameters are given in
  Table~\ref{para.tab}.  \label{pl_allSy.pdf}}
\end{figure}

\subsection{Fits to Seyfert nuclei}

\cite{GarciaBurillo14} presented ALMA continuum observations of
    NGC~1068 and find that $r_{\rm {out}}/r_{\rm {in}} \sim 100$.
    This is much larger than the value inferred by \cite{AH11}
    ($r_{\rm {out}}/r_{\rm {in}} \sim 6$ ) who used MIR
    data. \cite{Ichi15} also on the basis of NIR and MIR data, find
    that this ratio is less than 20 for a sample Seyferts.  It is
    expected and confirmed by our models that the ratio derived from
    submillimeter data, which is probing dust at $\sim 30$\,K or
    below, will be larger than the ratio derived from MIR data which
    is probing dust at 300\,K.

    \cite{Lira13} find that in 50\% of the Seyferts that are not well
    fit by the clumpy torus model of \cite{Nenkova08}, a
    NIR excess is observed.  In our models such a NIR emission is
    provided by the homogeneous disk component.

    \cite{Ramos14b} present high spatial resolution 1 -- 18\,$\mu$m
    data of a sample of nearby undisturbed Seyfert galaxies with low
    to moderate amounts of foreground extinction ($A_{\rm V} \simless
    5$\,mag).  We model four objects from their sample (the Seyfert 1s
    NGC~1365, NGC~4151 and the Seyfert 2s NGC~3081 and NGC~5643) to
    test the ability of our SED library to fit available nuclear NIR
    and MIR photometry and spectroscopy.  We take high spatial
    resolution MIR spectroscopic observations of NGC~1365 from
    \cite{Burtscher13}, NGC~3081 and NGC~4151 from \cite{Ramos14b},
    and NGC~5643 from \cite{H10}. Our results are shown in
    Fig.~\ref{pl_allSy.pdf} and the fit parameters are summarized in
    Table~\ref{para.tab}. The torus models clearly under-predict the
    far infrared data (plotted as upper limits in the figure) as the
    emission at these wavelengths is dominated by the host galaxy.

\subsection{Two color IRAC diagram}

\begin{figure}
\centering \includegraphics[scale=0.45]{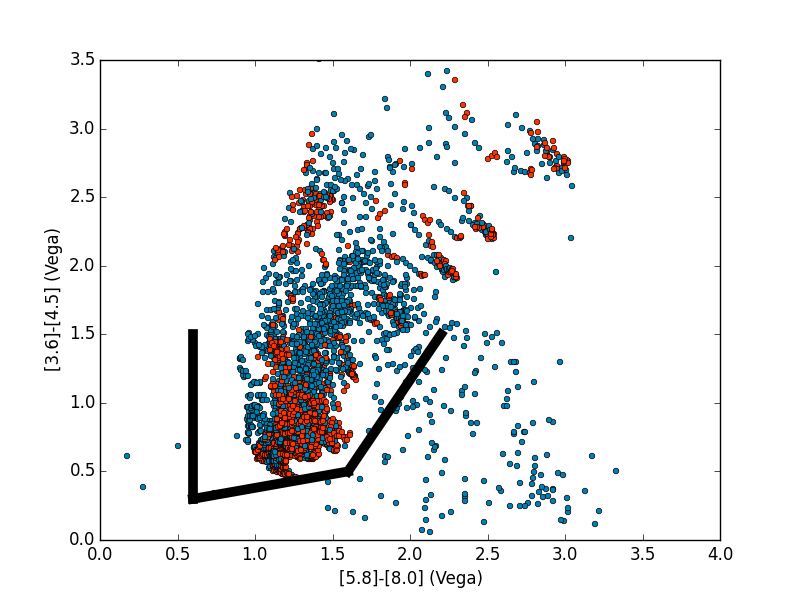}
\caption{Spitzer/IRAC mid IR color diagram.  Observed AGN populate
  the area in between the black lines at $[3.6] - [4.5] \simgreat 0.3$
  \citep{Stern05}. Colours derived from the AGN library populate this
  area. Type I sources with viewing angle $< 60^{\rm o}$ are shown as
  red circles and over-plotted are type II sources ($> 60^{\rm o}$) as
  blue circles \label{Stern}. }
\end{figure}

Mid IR photometry provides a simple technique for identifying active
galaxies.  \cite{Stern05} report on Spitzer mid IR colors, derived
from IRAC \citep{Fazio04}, of nearly 10,000 spectroscopically
identified AGN and galaxies. They find that simple mid IR color
criteria provide remarkably robust separation of AGN from normal
galaxies and stars. We indicate the empirical relation in
Fig.~\ref{Stern}. In that diagram AGN populate the area in between the
dashed lines at $[3.6] - [4.5] \simgreat 0.3$, whereas galaxies and
stars are below at $[3.6] - [4.5] \simless 0.3$ and $-0.3 <
[5.8]-[8.0] < 3.2$.

We compute the IRAC colours for each spectrum of the AGN
library. Further radiation components that may contribute to the IRAC
fluxes are ignored.  The IRAC magnitudes of the library are computed
using filter curves and flux conversion as described in the IRAC Data
Handbook. We find that most sources of the AGN library would be
classified as AGN when the empirical mid IR selection criteria by
\cite{Stern05} is applied. Although there are some additional sources
in the area near $[5.8]-[8.0] \simgreat 2$, these are edge--on sources
with high optical depth.

\section{Conclusion \label{conclusion.sec}}

We assume that dust in the AGN torus is distributed in a clumpy medium
or in a homogeneous disk or as a combination of the two (i.e. a
2-phase medium). We have computed with a self--consistent three
dimensional radiative transfer code the SED of such an AGN structure
over a wide range of their basic parameters: the viewing angle, the
inner radius, the volume filling factor and optical depth of the
clouds, and the optical depth of the disk midplane. We find that
details of the cloud properties, such as cloud size, geometrical
shape, or their density structure, as well as the presence of clumps
such as those detected by X--ray eclipse observations, have a minor
impact on the AGN dust emission spectrum.

We visualize the applied three-dimensional dust density structure of
the AGN. The AGN torus emits anisotropically and generally the SED
depends on the viewing direction. This happens for wavelengths below
the far IR peak flux. At longer wavelengths the AGN emission becomes
isotropic. We also show this in AGN images, where the scattering light
($\sim 1$\,$\mu$m) and the warm dust emission ($\sim 10$\,$\mu$m) is
anisotropic, while the cold dust emission ($\simgreat 100$\,$\mu$m)
appears isotropic. The precise wavelength where the AGN becomes
isotropic depends on the dust cross section. We find that type I
sources emit more IR radiation than type IIs.  When the disk material
is included there is additional hot dust in the system, which may
increase the near IR flux for face-on views by an order of magnitude.

AGN torus models often consider dust similar to what is assumed for
the diffuse ISM.  Because of the much higher density and stronger
radiation environment in the AGN we favour fluffy grains.  Their
absorption cross section is generally larger than for ISM dust
especially in the far IR and submillimeter being a factor 10 higher at
1\,mm.  We show that for unresolved observations, photon scattering
within the beam may increase the detected flux and alter the
wavelength dependence of the extinction curve. This implies that there
is no direct one--to--one link between the observed extinction curve
and the wavelength dependence of the dust cross sections. Claims of
detecting grain growth in AGN tori that are based on extinction
measurements should therefore be treated with caution.

The influence of the model parameters on the SED and their impact on
the strength of the 10\,$\mu$m silicate band is discussed. The AGN
library accounts well for the observed scatter of the 10\,$\mu$m feature
strengths and peak emission. The peak is centered near 10.4\,$\mu$m for
ISM dust, whereas it is shifted to even longer wavelengths ($\sim
11.5\,\mu$m) for fluffy grains, and this is more consistent with
observations. We cannot identify a striking need to postulate clumpy
AGN torus models when explaining observed properties of the silicate
band. However, arbitrary clump distributions of the AGN torus ease
accounting for silicate features observed at various strengths and
viewing angles.

In the UV, optical and near IR, it may be necessary to add to the
library SED a low luminosity component. This may be for type I sources
beamed synchrotron radiation or for type II sources stellar light from
the galactic disk.  These photons escape the torus without
interaction. Given a set of data points for a particular AGN, there is
a simple procedure to select from the library those elements which
best match them.  If the observations cover the full wavelength bands
from the near IR to submm, one usually finds a library element that
fits very well. This is demonstrated for a number of representative
type I and type II objects, the famous hyperluminous infrared galaxy
IRAS F10214+4724, and 4 nearby Seyferts.  In all cases, the parameters
defining the element are in full accord with what is known about the
object from other studies. Interestingly for the luminous AGN it is
possible to explain the SED of these objects with pure AGN models
without a need to postulate starburst activity. In a second step we
assume a combination of AGN and starburst activity with the goal of
estimating an upper limit of the starburst luminosity. We find that
the SED of the 5 high luminosity objects studied in this paper,
require a starburst luminosity of typically 15\% (with a range between
3 - 33\%) of that of the AGN. In our model the starburst contribution
to the far IR emission of the AGN can well be marginal.  We also
confirm that objects in the library span the range of mid IR colors as
empirically established from observations.

Our library of AGN models promises to be very useful for interpreting
the detailed post-{\em Spitzer} and post-{\em Herschel} SED of large
samples of AGN and (ultra-) luminous infrared galaxies in general as
well as high spatial resolution imaging observations with large
ground-based telescopes and interferometers. The library allows one to
constrain the most fundamental properties of the dust torus which is
powered by the central AGN. The SED library also provides a utility
that allows converting the observed ('apparent') AGN luminosity, which
is usually what is quoted in the literature, to the {\it {intrinsic}}
('true') luminosity of the AGN source.  The SEDs are accessible in a
public library \footnote{The SED library of the AGN models is
  available at: \\{\tt http://www.eso.org/\~\/rsiebenm/AGN/}}.


\begin{acknowledgements} {We are grateful to Endrik Kr\"ugel for
    helpful discussions and Bruno Altieri for supporting us in the
    analysis of the Herschel photometry. We also thank Cristina Ramos
    Almeida and Sebastian H\"onig for providing us their MIR spectra in
    electronic form. We are greatful to Seth Johnson for discussions
    and help concerning the Bayesian SED fitting technique. This
    research has made use of the NASA/IPAC Extragalactic Database
    (NED) which is operated by the Jet Propulsion Laboratory,
    California Institute of Technology, under contract with the
    National Aeronautics and Space Administration. We use the NASA/
    IPAC Infrared Science Archive, which is operated by the Jet
    Propulsion Laboratory, California Institute of Technology, under
    contract with the National Aeronautics and Space Administration. }
\end{acknowledgements}

\bibliographystyle{aa} 
\bibliography{References}

\end{document}